\documentclass[prd,aps,preprint,amsmath,amssymb,eshowkeys,nofootinbib]{revtex4-1}
\pdfoutput=1
\usepackage{verbatim,graphics,graphicx,color,slashed,textcomp,bbm,mathdots}
\usepackage[normalem]{ulem} 
\usepackage[colorlinks=true,linkcolor=red,urlcolor=blue,citecolor=blue]{hyperref}

\graphicspath{{figures/}}

\usepackage[utf8]{inputenc}

\begin{document}

\title{Dark matter production in Weyl $R^2$ inflation}

\author{Qing-Yang Wang$^{a}$}
\author{Yong Tang$^{a,b,c,d}$}
\author{Yue-Liang Wu$^{a,b,c,e}$}
\affiliation{\begin{footnotesize}
		${}^a$University of Chinese Academy of Sciences (UCAS), Beijing 100049, China\\
		${}^b$School of Fundamental Physics and Mathematical Sciences, \\
		Hangzhou Institute for Advanced Study, UCAS, Hangzhou 310024, China \\
		${}^c$International Center for Theoretical Physics Asia-Pacific, Beijing/Hangzhou, China \\
		${}^d$National Astronomical Observatories, Chinese Academy of Sciences, Beijing 100101, China\\
		${}^e$Institute of Theoretical Physics, Chinese Academy of Sciences, Beijing 100190, China
		\end{footnotesize}}
	
\date{\today}

\begin{abstract}
	Dark matter and inflation are two key elements to understand the origin of cosmic structures in modern cosmology, and yet their exact physical models remain largely uncertain. The Weyl scaling invariant theory of gravity may provide a feasible scheme to solve these two puzzles jointly, which contains a massive gauge boson playing the role of dark matter candidate, and allows the quadratic scalar curvature term, namely $R^2$, to realize a viable inflationary mechanism in agreement with current observations. We ponder on the production of dark matter in the Weyl $R^2$ model, including the contribution from the non-perturbative production due to the quantum fluctuations from inflationary vacuum and perturbative ones from scattering. We demonstrate that there are generally three parameter regions for viable dark matter production: (1) If the reheating temperature is larger than $10^3~\mathrm{GeV}$, the Weyl gauge boson as dark matter can be produced abundantly with mass larger than the inflation scale $\sim 10^{13}~\mathrm{GeV}$. (2) Small mass region with $3\times10^{-13}~\mathrm{GeV}$ for a higher reheating temperature. (3) Annihilation channel becomes important in the case of higher reheating temperature, which enables the Weyl gauge boson with mass up to $4\times10^{16}~\mathrm{GeV}$ to be produced through freeze-in.
	
\end{abstract}

\maketitle
\newpage

\section{Introduction}

Dark matter is an effective paradigm to explain the various observations in galactic rotation curve, cosmic large scale structures, and the gravitational lensing of clusters \cite{grc1,lss1,lensing1,lensing2}. Studies of the primordial nucleosynthesis \cite{nucl} and the power spectrum of cosmic microwave background (CMB) \cite{Planck2018} also provide strong evidence that there is abundant missing matter whose present total amount is about five times that of visible matter described by the standard model of particle physics (SM). As for the physical essence of dark matter, however, it still remains obscure. All experiments so far have not identified whether dark matter interacts with SM particles except gravity \cite{detect1,detect2,detect3}.

In this paper, we ponder on a new kind of vector dark matter~\cite{WGB1,WGB2} from the Weyl scaling invariant theory of gravity~\cite{Wu:2015wwa, Wu:2017urh}. Weyl symmetry with various forms has wide applications in cosmological physics~\cite{Weyl1,Weyl3,Weyl4,Weyl5,Weyl6}. It can naturally introduce a local scaling symmetry associated with a gauge boson (called Weyl gauge boson or WGB), which acquires a mass from the symmetry breaking. The consideration of Weyl gravity is also motivated by the reasonableness of introducing a quadratic scalar curvature term $R^2$ into theory~\cite{R2Weyl2,R2Weyl3,R2Weyl4}. It can provide a successful inflationary mechanism with observational conformity, albeit it has a different form under the Weyl geometry \cite{R2Weyl1,R2Weyl5,R2Weyl6,R2Weyl7}. Our previous study has discussed Weyl $R^2$ inflation \cite{R2Weyl4} and briefly the production of dark matter. However, the intrinsic connection between the inflaton and WGB has not been fully explored. The purposes of this paper are twofold. First, we use the latest cosmological observations to update the preferred parameter space for Weyl $R^2$ inflation. Second, we identify the viable dark matter mass range by considering all the production channels. 

There are several contributions for the production of WGB. If dark matter is already massive during inflation, there will be an important non-perturbative production mechanism~\cite{Parker1,Parker2,Ford,Lyth,gpdm,nonthermal,vdm1,pgdm1,pgdm2,vdm2, Chung:2018ayg, Kolb:2020fwh} that is due to the quantum fluctuations in rapidly expanding background~\cite{Ema:2016dny, Ema:2016hlw, Li:2020xwr, Li:2019ves, Herring:2020cah, Babichev:2018mtd, Hashiba:2018tbu, Ling:2021zlj}. Other important contributions are also present after inflation, such as producing from inflaton decay or the annihilation from particles in thermal bath after reheating~\cite{Garny:2015sjg, Tang:2016vch, pgdm3, Garny:2017kha, Chen:2017kvz, Bernal:2018qlk, Aoki:2022dzd, Clery:2021bwz, Frangipane:2021rtf, Redi:2021ipn, Barman:2021qds, Mambrini:2021zpp, Haque:2021mab, Clery:2022wib}. The previous study \cite{R2Weyl4} has only estimated the annihilation channel through scalars, but not taken into account the gravitational annihilation, non-perturbative production and the decay process. Our goal here is to figure out how these production mechanisms determine the relic abundance of WGB as dark matter, and identify the possible mass ranges. 

The paper is organized as follows. In Sec.~II, we introduce the formalism of Weyl $R^2$ gravity with inflaton and massive WGB. Then in Sec.~III, we discuss the constraints on the Weyl $R^2$ inflation from the latest observations. Later in Sec.~IV, we focus on the WGB as dark matter. We first discuss the equation of motion and quantization of the Weyl gauge field in a cosmological background, and elaborate the method to calculate the energy density of WGB produced from the inflationary fluctuations. Then based on the inflation and reheating process, we compute the production rate and final relic abundance of WGB from all possible channels, and show the constraints on the mass and reheating temperature in detail. Finally, we summarize and give our conclusion.

\section{Formalism and model}

We use the following conventions in this paper: the metric $\eta_{\mu\nu}=(-1,+1,+1,+1)$, and natural unit $\hbar=c=1$, $M_P\equiv1/\sqrt{8\pi G}=2.435\times10^{18}~\mathrm{GeV}=1$.

The complete Lagrangian that is motivated by gauge theory of gravity can be found in Refs.~\cite{Wu:2015wwa, Wu:2017urh, Wu:2021ign, Wu:2021ucc}. For the interest of dark matter production and inflation, we can use the metric formalism and illustrate by starting with the following simplest global scaling-invariant Lagrangian $\mathcal{L}$ with linear and quadratic scalar curvature
\begin{equation}
	\frac{\mathcal L}{\sqrt{-g}}= \frac{1}{2}\phi^2 R+\frac{\beta}{12} R^2-\frac{\lambda}{2} \partial^\mu\phi \partial_\mu\phi ,
	\label{L0}
\end{equation}
where $g$ is the determinant of metric tensor $g_{\mu\nu}$, Ricci scalar $R$ are defined as usual and $\lambda$ is a nonzero real number. A scalar field $\phi$ is introduced here to maintain the global scaling symmetry with the following transformation rules,
\begin{equation}
	\begin{aligned}
		\mathrm{metric}:~&g_{\mu\nu}\rightarrow g'_{\mu\nu}=f^2g_{\mu\nu},\\
		\mathrm{scalar}:~&\phi\rightarrow \phi'=f^{-1}\phi,\\
		\mathrm{Ricci~scalar}:~&R\rightarrow R'=f^{-2}R,
		\label{CT}
	\end{aligned}
\end{equation}
where $f$ is a constant scale factor. Note that one can introduce a scalar potential $\phi^4$ without spoiling the symmetry, but for our purpose in this paper, such a term is not necessary and we neglect here. 

For the global scaling transformation, the Christoffel connection
\begin{equation}
	\Gamma^\rho_{\mu\nu}=\frac{1}{2}g^{\rho\sigma}(\partial_\mu g_{\sigma\nu}+\partial_\nu g_{\mu\sigma}-\partial_\sigma g_{\mu\nu}),
	\label{Gamma}
\end{equation}
is invariant. When we extend the global symmetry into the local one, namely $f\rightarrow f(x)$, the connection and $\mathcal{L}$ would not be invariant. To make the theory locally scaling invariant, we construct the following connection
\begin{equation}
	\hat{\Gamma}^\rho_{\mu\nu}=\Gamma^\rho_{\mu\nu}+(W_\mu g^\rho_\nu+W_\nu g^\rho_\mu-W^\rho g_{\mu\nu}),
	\label{WeylGamma}
\end{equation}
by introducing a vector $W_\mu\equiv g_W w_\mu$, called Weyl gauge field ($g_W$ is the gauge coupling), with the following transformation rule,
\begin{equation}
	W_\mu\rightarrow W'_\mu=W_\mu-\partial_\mu \ln f(x).
	\label{CTW}
\end{equation}
$\hat{\Gamma}^\rho_{\mu\nu}$ can be also obtained by replacing $\partial_\mu g_{\rho \sigma}\rightarrow (\partial_\mu +2 W_\mu)g_{\rho \sigma}$ in Eq.~\ref{Gamma}. Then the corresponding Ricci scalar $\hat{R}$ should be defined through $\hat{\Gamma}^\rho_{\mu\nu}$ 
and Lagrangian in Eq.~(\ref{L0}) is modified to a new form
\begin{equation}
	\frac{\mathcal L}{\sqrt{-g}}=\frac{1}{2}\phi^2\hat R+\frac{\beta}{12}\hat R^2-\frac{\lambda}{2} D^\mu\phi D_\mu\phi-\frac{1}{4g_W^2}F_{\mu\nu}F^{\mu\nu},
	\label{L1}
\end{equation}
where the derivative in the kinetic term of $\phi$ has been replaced to the covariant form, $D_\mu=\partial_\mu-W_\mu$, and the invariant field strength of $W_\mu$ is defined as $F_{\mu\nu}\equiv\partial_\mu W_\nu-\partial_\nu W_\mu$. This model is also discussed in \cite{R2Weyl2,R2Weyl3,R2Weyl4}. 

To make the theoretical formalism more transparent for analyzing inflation and dark matter, we shall utilize the equivalence between $f(R)$ theory and the scalar-tensor theory~\cite{fR1,fR2}. We define
\begin{equation}
	F(\hat R)=\frac{1}{2}\phi^2\hat R+\frac{\beta}{12}\hat R^2,
\end{equation}
and introduce an auxiliary scalar field $\chi$. Then Eq.~(\ref{L1}) can be written as an equivalent form
\begin{equation}
	\begin{aligned}
		\frac{\mathcal L}{\sqrt{-g}}=&\frac{1}{2}[F'(\chi^2)(\hat R-\chi^2)+F(\chi^2)]-\frac{\lambda}{2}D^\mu\phi D_\mu\phi-\frac{1}{4g_W^2}F_{\mu\nu}F^{\mu\nu} \\
		=&\frac{1}{2}\left(\phi^2+\frac{\beta}{3}\chi^2\right)\hat R-\frac{1}{12}\beta\chi^4-\frac{\lambda}{2}D^\mu\phi D_\mu\phi-\frac{1}{4g_W^2}F_{\mu\nu}F^{\mu\nu},
		\label{L2}
	\end{aligned}
\end{equation}
where $F'(\chi^2)\equiv \left|\frac{\partial F(\hat R)}{\partial \hat R}\right|_{\hat R=\chi^2}$. The equivalence relation $\chi^2=\hat R$ can be easily derived by using the equation of motion from $\delta\mathcal L/\delta\chi=0$.

Note that the local scaling symmetry of Eq.~(\ref{L2}) allows us to choose convenient $f(x)$ in Eq.~(\ref{CT}) and (\ref{CTW}) to fix the gauge and simplify the Lagrangian. To compare with the Einstein gravity, we adopt $f(x)=\sqrt{\phi^2+\frac{\beta}{3}\chi^2}$ and get $\frac{1}{2}(\phi^2+\frac{\beta}{3}\chi^2)\hat R=\frac{1}{2}\hat R$, or equivalently, $\phi^2+\frac{\beta}{3}\chi^2=1$.
Putting things together, we can obtain the following Lagrangian
\begin{equation}
	\begin{aligned}
		\frac{\mathcal L}{\sqrt{-g}}=&\frac{1}{2}R-\frac{\lambda}{2}D^\mu\phi D_\mu\phi-\frac{3}{4\beta}(1-\phi^2)^2-\frac{1}{4g_W^2}F_{\mu\nu}F^{\mu\nu}-3W^\mu W_\mu\\
		=&\frac{1}{2}R-\frac{3\lambda}{6+\lambda\phi^2}\partial^\mu\phi\partial_\mu\phi-\frac{3}{4\beta}(1-\phi^2)^2-\frac{1}{4g_W^2}F_{\mu\nu}F^{\mu\nu}\\
	    &-\frac{1}{2}(6+\lambda\phi^2)\left[W_\mu-\frac{1}{2}\partial_\mu\ln|6+\lambda\phi^2|\right]^2.
		\label{L3}
	\end{aligned}
\end{equation}
where we have used the relation between $\hat{R}$ and $R$,
\begin{equation}
	\hat R=R-6W_\mu W^\mu-\frac{6}{\sqrt{-g}}\partial_\mu(\sqrt{-g}W^\mu),
	\label{WeylR}
\end{equation}
and neglect the surface term at infinite due to the total derivative term above. 

We define a new field variable $\Phi$,
\begin{equation}
	\Phi\equiv \pm
	\begin{cases}
		\sqrt 6 \ln\left[\sqrt{\frac{+\lambda\phi^2}{6}}+\sqrt{\frac{+\lambda\phi^2+6}{6}}\right]~\mathrm{for}~ \lambda>0,\\
		\sqrt 6 \ln\left[\sqrt{\frac{-\lambda\phi^2}{6}}+\sqrt{\frac{-\lambda\phi^2-6}{6}}\right]~\mathrm{for}~ \lambda<0,
	\end{cases}
\end{equation}
to get the canonical kinetic term, and make transformation for Weyl gauge field
\begin{equation}
	\tilde W_\mu\equiv W_\mu-\frac{1}{2}\partial_\mu\ln|6+\lambda\phi^2|\equiv g_W\tilde w_\mu.
\end{equation}
Then the Lagrangian of Weyl $R^2$ model can be rewritten as a simple form
\begin{equation}
	\frac{\mathcal L}{\sqrt{-g}}=\frac{1}{2}R-\frac{1}{2}\partial^\mu\Phi\partial_\mu\Phi-\frac{1}{4g_W^2}\tilde F_{\mu\nu}\tilde F^{\mu\nu}-\frac{1}{2}m^2(\Phi)\tilde W^\mu\tilde W_\mu-V(\Phi),
	\label{L4}
\end{equation}
where we have defined the mass term
\begin{equation}
	m^2(\Phi)=
	\begin{cases}
		+6\cosh^2\left(\frac{\Phi}{\sqrt{6}}\right) ~\mathrm{for}~ \lambda>0,\\
		-6\sinh^2\left(\frac{\Phi}{\sqrt{6}}\right) ~\mathrm{for}~ \lambda<0,\\
	\end{cases}
	\label{m2}
\end{equation}
and scalar potential
\begin{equation}
	V(\Phi)=
	\begin{cases}
		\frac{3}{4\beta}\left[1-\frac{6}{\lambda}\sinh^2\left(\frac{\Phi}{\sqrt{6}}\right)\right]^2 ~\mathrm{for}~ \lambda>0,\\
		\frac{3}{4\beta}\left[1+\frac{6}{\lambda}\cosh^2\left(\frac{\Phi}{\sqrt{6}}\right)\right]^2 ~\mathrm{for}~ \lambda<0.
	\end{cases}
	\label{V}
\end{equation}

Now it is clear that this theory can be reformulated as the Einstein gravity with a massive Weyl gauge boson $\tilde W_\mu$ (WGB) and a self-interacting scalar $\Phi$. In the following sections, we shall show that $\Phi$ can be the inflaton field while $\tilde W_\mu$ is a dark matter candidate due to the $Z_2$ symmetry, $\tilde W_\mu\rightarrow - \tilde W_\mu$.

\section{Inflation in Weyl $R^2$ Gravity}

Now we concentrate on the inflationary mechanism in this model. The physics of $R^2$ term has been extensively studied as the well-known Starobinsky model \cite{Staro1,Staro2,Staro3,Staro4,Staro5,Staro6,Staro7,Staro8}. It was proposed originally to solve the cosmological initial singularity problem, and later demonstrated to be an elegant mechanism to give rise to an inflationary universe. In this section, we shall show the predictions in our model are different from those in Starobinsky model. 

\begin{figure}
	\centering
	\includegraphics[width=0.75\textwidth]{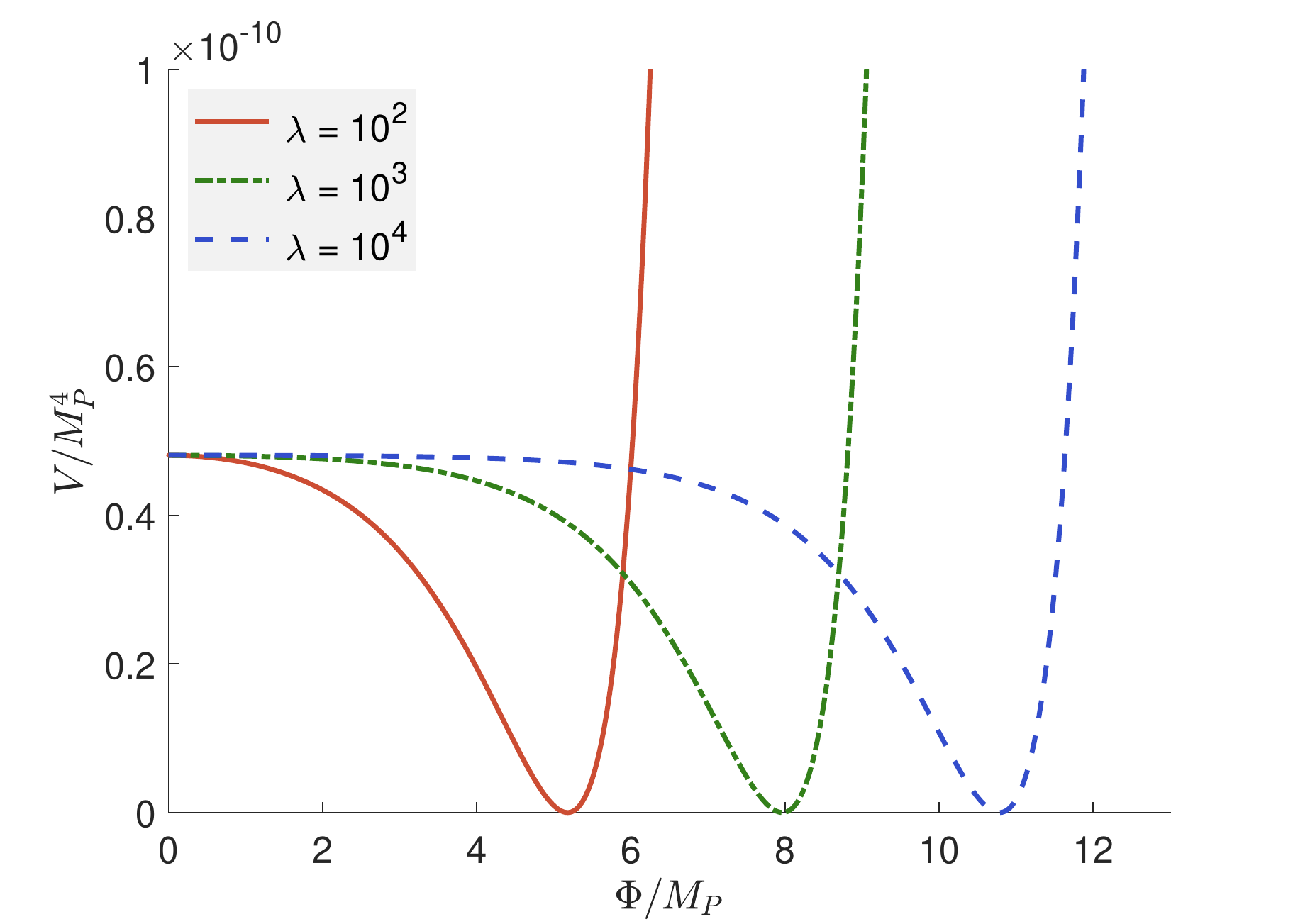}
	\caption{The inflationary potentials of the Weyl $R^2$ model with $\lambda=10^2,10^3,10^4$. It is a hilltop-like potential whose typical vacumm expectation value is larger than the Planck scale.}
	\label{Vphi}
\end{figure}

The potential $V(\Phi)$ in Fig.~\ref{Vphi} for a hilltop-like shape regardless of the sign of $\lambda$. The parameter $\beta$ decides the height of the hilltop, and $\lambda$ determines the position of valley. Fig.~\ref{Vphi} shows the potentials with $\beta=1.56\times10^{10}$ in Planck unit and several values of $\lambda>0$ ($\lambda<0$ is similar because $\cosh^2x=1+\sinh^2x$).

At the potential minimum $\Phi$ can outstrip the Planck scale when $|\lambda|$ is not too small. It is a general picture that for large $|\lambda|$ the potential can be flat and lead to an inflationary universe when $\Phi$ slow-rolls on the hilltop, until it falls into the valley. 

\begin{figure}
	\centering
	\includegraphics[width=0.75\textwidth]{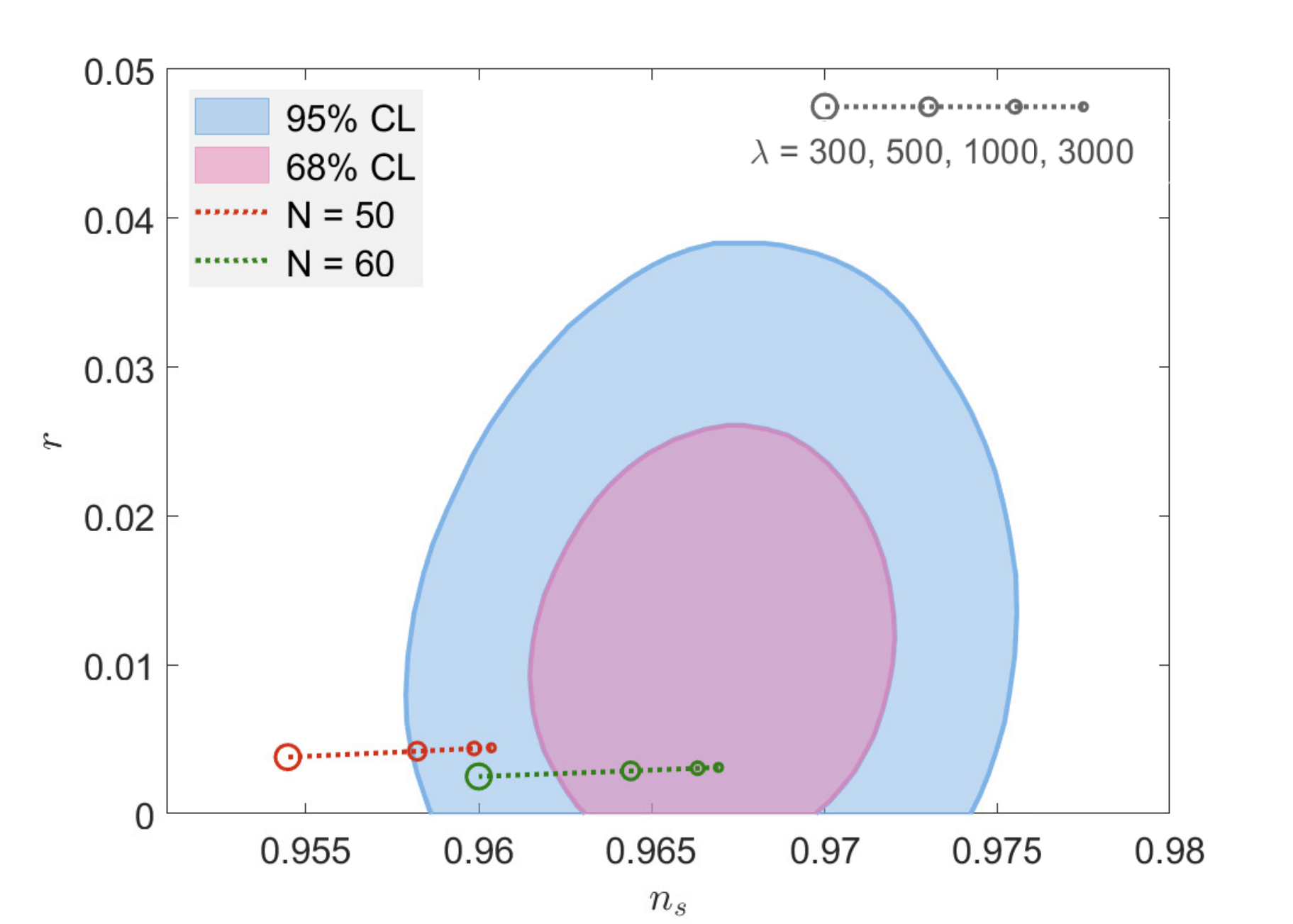}
	\caption{The predictions of spectral index $n_s$ and tensor-to-scalar ratio $r$ for the Weyl $R^2$ model with $\lambda$ from 300 to 3000, and e-folding number $N=50,60$, in comparison with the recent constraints shown by the BICEP/$Keck$ Collaboration \cite{BICEP}. }
	\label{r_ns}
\end{figure}

To compare with the latest observational constraints by BICEP/$Keck$ Collaboration \cite{BICEP}, we calculate the spectral index $n_s$ and tensor-to-scalar ratio $r$ for the primordial perturbations. These values are related to the slow-roll parameters $\epsilon$ and $\eta$ through
\begin{equation}
	\begin{aligned}
		&\epsilon=\frac{1}{2}\left[\frac{V'(\Phi)}{V}\right]^2,~\eta=\frac{V''(\Phi)}{V},~n_s=1-6\epsilon+2\eta,~r=16\epsilon.
	\end{aligned}
\end{equation}
We numerically compute $n_s$ and $r$ for this model with $\lambda$ from 300 to 3000, and show the results in Fig.~\ref{r_ns}, where $N$ is defined as the e-folding number of the inflationary multiples from the horizon scale as large as the present one to the end of inflation \cite{efolds}, $N\equiv\ln(a_e/a_i)\sim(50,60)$. The figure indicates that, as $\lambda$ increases, $n_s$ and $r$ will approach fixed values predicted in Starobinsky model for every e-folds number $N$. The physical reasons has been explained in detail in Ref.~\cite{R2Weyl4} in which the authors show the equivalence between this model and Starobinsky inflation as $|\lambda|\rightarrow +\infty$. There is another noteworthy point in Fig.~\ref{r_ns}, the typical tensor-to-scalar ratio in this model is greater than $\mathcal O(10^{-3})$, which implies that the Weyl $R^2$ model may be tested by the next generation experiment of CMB B-mode polarization \cite{CMB-S4}. Current observation can give constraints on $\lambda\gtrsim 500$ for $N=50$ and $\lambda \gtrsim 250$ for $N=60$. 

On the other hand, the observation of CMB perturbation also provide a constraint on the Hubble parameter during inflation (denoted by $H_{\mathrm{inf}}$) and the mass of inflaton $m_\Phi$ in this model. It is known that the amplitude of scalar spectra $\Delta^2_s\sim V/24\pi^2\epsilon$, and the observation of CMB has confirmed that $\Delta^2_s\sim 2.1\times10^{-9}$. Thus this result constrains the parameter $\beta$ in Eq.~(\ref{V}), which determines the height of the potential $V$ at the hilltop. In this way, we find the Hubble parameter during inflation can be limited to $H_{\mathrm{inf}}\sim\sqrt{V/3}\sim10^{13}~\mathrm{GeV}$. Similarly, the mass of inflaton is also constrained to $m_\Phi\sim2\times10^{13}~\mathrm{GeV}$ defined at the minimum of the scalar potential.

\section{Weyl gauge boson as dark matter}

Now we discuss the physics of Weyl gauge boson. As we have shown, the Lagrangian has $Z_2$ symmetry, $\tilde W_\mu\rightarrow - \tilde W_\mu$, which enables the stability of $\tilde W_\mu $. The mass term $-\frac{1}{2}m^2\tilde W^\mu\tilde W_\mu$ in Eq.~(\ref{L4}), however, shows that negative sign would appear for $\lambda<0$. This does not necessarily mean an inconsistent theory because one could introduce another scalar field $\psi$ with $D_\mu \psi D^\mu \psi$ that can contribute to the mass of $\tilde W_\mu$ when $\psi \neq 0$ at its potential minimum. For illustration, we shall only focus on the $\lambda >0$ case in the rest of this paper. 

Since the magnitude of $\Phi$ is evolving in the inflation and subsequent reheating process, the mass term in Eq.~(\ref{m2}) is also changing in this period, until $\Phi$ completely stops at its potential minimum $\Phi_0$. We can expand the mass term at $\Phi=\Phi_0$ as
\begin{equation}
	\begin{aligned}
		m^2=&6\cosh^2\left(\frac{\varphi+\Phi_0}{\sqrt{6}}\right)\\
		=&6\cosh^2\left(\frac{\Phi_0}{\sqrt6}\right)+\sqrt6\sinh\left(\frac{2\Phi_0}{\sqrt6}\right)\varphi+\cosh^2\left(\frac{\Phi_0}{\sqrt6}\right)\varphi^2+\mathcal O(\varphi^3).
		\label{m2expand}
	\end{aligned}
\end{equation}
It implies that the term $-\frac{1}{2}m^2\tilde W^\mu\tilde W_\mu$ not only describes the mass of WGB, but also $\varphi-\tilde W$ interactions that may lead to the decay or annihilation of inflaton into WGB after inflation. The eventual mass of WGB is determined by the parameters $\lambda$ and $g_W$
\begin{equation}
	m_W=\sqrt{6}|g_W|\cosh[\Phi_0(\lambda)/\sqrt6].
	\label{mW}
\end{equation}
The couplings between $\varphi$ and WGB also depend on these two parameters and can be read from above. In the rest of this section, we shall discuss the production mechanism for WGB as a dark matter candidate in detail by including all of the above couplings and interactions. 

\subsection{Weyl gauge field in a cosmological background}
The production of vector dark matter includes both contributions from perturbative scattering and non-perturbative mechanism from vacuum fluctuations due to the quantum effect in curved background~\cite{vdm1,pgdm1, pgdm2, vdm2}. Let us concentrate on the part of WGB in Eq.~(\ref{L4})
\begin{equation}
	S=\int d^4x\sqrt{-g}\left[-\frac{1}{4g_W^2}\tilde F_{\mu\nu}\tilde F^{\mu\nu}-\frac{1}{2}m^2\tilde W^\mu\tilde W_\mu\right].
	\label{SW}
\end{equation}
The metric is set to the FLRW form
\begin{equation}
	g_{\mu\nu}dx^\mu dx^\mu=a^2(\tau)(-d\tau^2+d\vec x^2),
\end{equation}
where $a$ is the scale factor of the universe, and $\tau$ is the conformal time, which has the following relation with the physical time $d\tau=dt/a$. We can rewrite Eq.~(\ref{SW}) in components
\begin{equation}
	S=\int d^3xd\tau\frac{1}{2}\left(|\vec{\tilde w}'-\nabla\tilde w_0|^2-|\nabla\times\vec{\tilde w}|^2+a^2\tilde m^2\tilde w_0^2-a^2\tilde m^2|\vec{\tilde w}|^2\right),
	\label{SWcomp}
\end{equation}
where $'\equiv\frac{\partial}{\partial\tau}$, $\tilde m\equiv|g_W|m(\Phi)$, and $\vec{\tilde w}$ denotes the spatial components.

It is obvious that $\tilde w_0$ in Eq.~(\ref{SWcomp}) has no kinetic term, which is commonly known as that only three of the four components of massive vectors are independent  and physical modes. Hence, we can decouple the part of $W_0$ from Eq.~(\ref{SWcomp}) with Fourier transform, and only retain the physical modes, including two transverse modes and a longitudinal mode
\begin{equation}
	\vec{\tilde w}(\tau,\vec k)\equiv\sum_{j=\pm,L}\vec\epsilon_j(\vec k)W_j(\tau,\vec k),
\end{equation}
where $\vec\epsilon_j(\vec k)$ is the orthonormal polarization vector. Their actions are derived as follows
\begin{equation}
	S_\pm=\int\frac{d^3kd\tau}{(2\pi)^3}\frac{1}{2}\left(|W_\pm'|^2-(k^2+a^2\tilde m^2)|W_\pm|^2\right),
	\label{SWT}
\end{equation}
\begin{equation}
	S_L=\int\frac{d^3kd\tau}{(2\pi)^3}\frac{1}{2}\left(\frac{a^2\tilde m^2}{k^2+a^2\tilde m^2}|W_L'|^2-a^2\tilde m^2|W_L|^2\right),
	\label{SWL}
\end{equation}
where $k$ is the magnitude of comoving momentum $\vec k$, and $\vec k\cdot\vec{\tilde w}=kW_L$ is used in the derivation.

For the sake of simplicity, we redefine a new field $\mathcal W_L$ for the longitudinal mode
\begin{equation}
	\mathcal W_L\equiv\frac{a\tilde m}{\sqrt{k^2+a^2\tilde m^2}}W_L.
\end{equation}
Then the equation of motion for each physical mode can be written as oscillator equations
\begin{equation}
	W_\pm''+\omega_\pm^2W_\pm=0,
	\label{oEqWT}
\end{equation}
\begin{equation}
	\mathcal W_L''+\omega_L^2\mathcal W_L=0,
	\label{oEqWL}
\end{equation}
with frequency variables
\begin{equation}
	\omega_\pm^2=k^2+a^2\tilde m^2,
	\label{w2WT}
\end{equation}
\begin{equation}
	\begin{aligned}
	&\omega_L^2=k^2+a^2\tilde m^2-\frac{k^2}{k^2+a^2\tilde m^2}\frac{a''}{a}+3\frac{k^2\tilde m^2a'^2}{(k^2+a^2\tilde m^2)^2}\\
	&-\frac{k^2}{k^2+a^2\tilde m^2}\frac{\tilde m''}{\tilde m}+3\frac{k^2a^2\tilde m'^2}{(k^2+a^2\tilde m^2)^2}-\frac{a'}{a}\frac{\tilde m'}{\tilde m}\frac{2k^2(k^2-2a^2\tilde m^2)}{(k^2+a^2\tilde m^2)^2}.
	\label{w2WL}
	\end{aligned}
\end{equation}
The mode functions of WGB can be determined by solving Eq.~(\ref{oEqWT}) and (\ref{oEqWL}). 

Next, we show the relation between the energy density and the mode functions. The energy-momentum tensor of field in a curved background is
\begin{equation}
	T_{\mu\nu}=\frac{-2}{\sqrt{-g}}\frac{\delta(\sqrt{-g}S)}{\delta g^{\mu\nu}},
	\label{Tuv}
\end{equation}
and the energy density is defined as the vacuum expectation of $T_{00}$
\begin{equation}
	\rho=\langle 0|T_{00}|0\rangle.
\end{equation}
Therefore, to calculate the energy density of Weyl gauge field, we need to define the vacuum state $|0\rangle$ of the theory. In other words, a quantized theory is required. The first step of quantization is to give the Heisenberg representation for Weyl gauge field
\begin{equation}
	\begin{aligned}
		\hat W_\pm(\tau,\vec x)=\int\frac{d^3k}{(2\pi)^{3/2}}&\left[\hat a_{\vec k}\vec\epsilon_\pm(\vec k)W_\pm(\tau,\vec k) e^{i\vec k\cdot\vec x}+\hat a_{\vec k}^\dag\vec\epsilon_\pm^*(\vec k)W_\pm^*(\tau,\vec k)e^{-i\vec k\cdot\vec x}\right],\\
		\hat{\mathcal W}_L(\tau,\vec x)=\int\frac{d^3k}{(2\pi)^{3/2}}&\left[\hat b_{\vec k}\vec\epsilon_L(\vec k)\mathcal W_L(\tau,\vec k) e^{i\vec k\cdot\vec x}+\hat b_{\vec k}^\dag\vec\epsilon_L^*(\vec k)\mathcal W_L^*(\tau,\vec k)e^{-i\vec k\cdot\vec x}\right].
		\label{Fmode}
	\end{aligned}
\end{equation}
Then the quantization demands that the ladder operators should satisfy the following commutative relations
\begin{equation}
	\left[\hat a_{\vec k},\hat a_{\vec k}^\dag\right]=\delta_{jj'}\delta(\vec k-\vec k'),~\left[\hat b_{\vec k},\hat b_{\vec k}^\dag\right]=\delta(\vec k-\vec k'),
	\label{ladder}
\end{equation}
where $j$ is the index that refers to the two components of transverse mode. The commutative relations can also be represented as a canonical form
\begin{equation}
	\begin{aligned}
		\left[\hat W_\pm(\tau,\vec x),\hat W_\pm'(\tau,\vec y)\right]=&i\delta(\vec x-\vec y),\\
		\left[\hat{\mathcal W}_L(\tau,\vec x),\hat{\mathcal W}_L'(\tau,\vec y)\right]=&i\delta(\vec x-\vec y),
	\end{aligned}
\end{equation}
which imply the following normalization condition
\begin{equation}
	W_\pm W_\pm'^*-W_\pm^*W_\pm'=\mathcal W_L\mathcal W_L'^*-\mathcal W_L^*\mathcal W_L'=i.
\end{equation}

Now we can define the vacuum $|0\rangle$ as a state that complies with $\hat a_{\vec k}|0\rangle=\hat b_{\vec k}|0\rangle=0$. Then the non-perturbative production of particles from vacuum can be calculated by $\rho=\langle 0|T_{00}|0\rangle$, described by the time-dependent mode functions. Finally, with Eq.~(\ref{SWT}), (\ref{SWL}), and (\ref{Tuv})-(\ref{ladder}), the energy density of Weyl gauge field in a cosmological background is obtained as
\begin{equation}
	\rho_\pm=\int\frac{k^2dk}{4\pi^2a^4}\left[|W'_\pm|^2+(k^2+a^2\tilde m^2)|W_\pm|^2-\omega_\pm\right],
	\label{rhoT}
\end{equation}
\begin{equation}
	\begin{aligned}
		\rho_L=\int\frac{k^2dk}{4\pi^2a^4}\bigg[&|\mathcal W'_L|^2+(k^2+a^2\tilde m^2)|\mathcal W_L|^2+\bigg(\frac{(\frac{a'}{a}+\frac{\tilde m'}{\tilde m})k^2}{k^2+a^2\tilde m^2}\bigg)^2|\mathcal W_L|^2\\
		&-\frac{(\frac{a'}{a}+\frac{\tilde m'}{\tilde m})k^2}{k^2+a^2\tilde m^2}(\mathcal W'_L\mathcal W^*_L+\mathcal W'^*_L\mathcal W_L)-\omega_L\bigg],
		\label{rhoL}
	\end{aligned}
\end{equation}
where the zero-point energy at flat spacetime has been subtracted by the term $-\omega$. The total energy density is the sum of two transverse modes and one longitudinal mode, $\rho_{tot}=2\rho_\pm+\rho_L$. 

Next we numerically solve the equations of motion (\ref{oEqWT}), (\ref{oEqWL}) in the inflationary background and calculate the energy density of WGB. Directly solving these equations might be inaccurate in some cases. For instance, when $\rho$ is many orders of magnitude smaller than the zero-point energy, the numerical calculation of $\rho$ will meet the situation that subtracting two very close numbers, which will make the numerical result unstable. Fortunately, this intractable case usually has an adiabatic vacuum, which means the vacuum evolves slowly. Then, a strategy called adiabatic approximation can be used to deal with the problems~\cite{reh1}. In the adiabatic approximation, the mode functions can be written as the following form
\begin{equation}
	\begin{aligned}
		&W_\pm(\tau)=\mathcal A_\pm^k(\tau) v_\pm^k(\tau)+\mathcal B_\pm^k(\tau) v_\pm^{k*}(\tau),\\
		&\mathcal W_L(\tau)=\mathcal A_L^k(\tau) v_L^k(\tau)+\mathcal B_L^k(\tau) v_L^{k*}(\tau),\\
		&v_{\pm(L)}^k=\frac{1}{\sqrt{2\omega_{\pm(L)}}}\exp\left[-i\int^\tau\omega_{\pm(L)}d\tau\right],
		\label{WABv}
	\end{aligned}
\end{equation}
where $\mathcal A^k(\tau)$ and $\mathcal B^k(\tau)$ are some functions for the $k$-mode, and satisfy the normalization condition
\begin{equation}
	|\mathcal A^k|^2-|\mathcal B^k|^2=1.
\end{equation}
Substituting Eq.~(\ref{WABv}) into Eq.~(\ref{Fmode}), we can redefine a new set of ladder operators by the Bogolyubov transformation
\begin{equation}
	\tilde{a}_{\vec k}=\hat a_{\vec k}\mathcal A^k+\hat a^\dag_{\vec k}\mathcal B^{k*},~ \tilde{b}_{\vec k}=\hat b_{\vec k}\mathcal A^k+\hat b^\dag_{\vec k}\mathcal B^{k*},
\end{equation}
which satisfy the similar commutative relations as Eq.~(\ref{ladder})
\begin{equation}
	\left[\tilde a_{\vec k},\tilde a_{\vec k}^\dag\right]=\delta_{jj'}\delta(\vec k-\vec k'),~\left[\tilde b_{\vec k},\tilde b_{\vec k}^\dag\right]=\delta(\vec k-\vec k').
	\label{ladder2}
\end{equation}
Then the energy density of the Weyl gauge field can be written as
\begin{equation}
	\rho_{\pm/L}=\int\frac{k^2dk}{2\pi^2a^4}|\mathcal B^k_{\pm/L}|^2\omega_{\pm/L}.
	\label{rhoB}
\end{equation}
Comparing to Eq.~(\ref{rhoT}) and (\ref{rhoL}), it can be seen the zero-point energy has disappeared in this expression, which means the aforementioned numerical problem is alleviated with the adiabatic approximation. In addition, the function $\mathcal B^k$ obeys
\begin{equation}
	\mathcal A' v=\frac{\omega_k'}{2\omega_k}v^*\mathcal B,~\mathcal B' v^*=\frac{\omega_k'}{2\omega_k}v\mathcal A,
\end{equation}
which are identical to Eq.~(\ref{oEqWT}) and (\ref{oEqWL}). 

Finally, it is essential to note that the adiabatic approximation can only be used to the case with positive $\omega^2$. This is the prerequisite for the existence of adiabatic vacuum. For the transverse mode Eq.~(\ref{w2WT}), this condition is completely satisfied. But for the longitudinal mode Eq.~(\ref{w2WL}), the tachyonic situation ($\omega_L^2<0$) is possible, and it is common in the case of $\tilde m\ll H_{\mathrm{inf}}$. Consequently, if the tachyonic situation appears in the longitudinal mode, it would violate the adiabatic condition. Then the mode functions and energy density can only be solved with original formulas (Eq.~(\ref{oEqWL}) and (\ref{rhoL})).

\subsection{Production of Weyl gauge boson}

Now we are in a position to apply the above formalism in the cosmological background of inflation. It can be seen from Eq.~(\ref{oEqWT})-(\ref{w2WL}) that, to solve the mode functions we also need the evolution of scale factor $a$ over conformal time $\tau$, which is determined by the inflation and reheating in our model. 

\subsubsection{Inflation and reheating}
The evolution of cosmic scale $a$ is described by the Friedmann equation
\begin{equation}
	\mathcal H^2\equiv(\frac{a'}{a})^2=\frac{a^2\rho}{3M_P^2},
	\label{Feq}
\end{equation}
where $'\equiv\frac{\partial}{\partial\tau}$. To solve this equation, we need to discuss how the matter energy density evolving in our model. To be more specific, we need reheating process to transfer the energy of the inflaton $\Phi$ into other relativistic particles, for example through decay into SM particles \cite{reh1,reh2,reh3,reh4,reh5}. Here, instead we first consider a new scalar $\psi$ which couples to the inflaton and SM particles simultaneously. Then in this way, the inflaton can first decay into $\psi$ before eventually transfer its energy to the SM particles. To maintain the scaling symmetry, we illustrate with the following toy model,
\begin{equation}
	\begin{aligned}
	\mathcal L_{\mathrm{reh}}=&-g_\psi^2\phi^2\psi^2-\frac{1}{2}D^\mu \psi D_\mu \psi +\mathcal L(\psi, SM)\\
	=&-\frac{6g_\psi^2}{\lambda}\sinh^2\left(\frac{\Phi}{\sqrt 6}\right)\psi^2-\frac{1}{2}D^\mu \psi D_\mu \psi+\mathcal L(\psi, SM)\\
	\sim&-\frac{g_\psi^2}{\lambda}\Phi^2\psi^2-\frac{1}{2}D^\mu \psi D_\mu \psi+\mathcal L(\psi, SM),
	\label{reh}
	\end{aligned}
\end{equation}
where $\mathcal L(\psi,SM)$ denotes the coupling between $\psi$ and SM particles, which we do not specify here. Here we have neglected higher order interactions in the last step. We reiterate that the definition $\Phi\equiv\varphi+\Phi_0$ has been introduced in the previous section. Then the first term in Eq.~(\ref{reh}) can be expanded as
\begin{equation}
	\begin{aligned}
		-\frac{g_\psi^2}{\lambda}\Phi^2\psi^2&=-\frac{g_\psi^2}{\lambda}(\varphi+\Phi_0)^2\psi^2\\
		&=-\frac{g_\psi^2}{\lambda}\Phi_0^2\psi^2-2\frac{g_\psi^2}{\lambda}\Phi_0\varphi\psi^2-\frac{g_\psi^2}{\lambda}\varphi^2\psi^2.
		\label{Lphichi}
	\end{aligned}
\end{equation}
From left are the mass term of $\psi$, the three-scalar vertex, and the four-scalar vertex. The feasibility of $\varphi$'s decay into $\psi$ requires the mass of $\psi$ must be less than the half that of $\varphi$. Note that the mass of inflaton is limited to $m_\Phi\sim2\times10^{13}~\mathrm{GeV}$ in our model, thus it is clear that $m_\psi\equiv g_\psi\Phi_0/\sqrt{\lambda}<10^{13}~\mathrm{GeV}$. With a typical VEV $\Phi_0\sim10M_P$, the coupling is limited to $g_\psi/\sqrt{\lambda}\lesssim4\times10^{-7}$, which gives an upper limit to the reheating temperature $T_r$. Note that the major factor determining $T_r$ is the $\varphi\rightarrow\psi\psi$ decay process caused by the three-scalar vertex in Eq.~(\ref{Lphichi}), then the probability of inflaton decay is easily derived as
\begin{equation}
	\Gamma(\varphi\rightarrow\psi\psi)=\frac{4g_\psi^4\Phi_0^2}{8\pi\lambda^2}\frac{\sqrt{(m_\Phi/2)^2-m_\psi^2}}{m_\Phi^2}=\frac{g_\psi^2m_\psi^2}{2\pi\lambda}\frac{\sqrt{(m_\Phi/2)^2-m_\psi^2}}{m_\Phi^2}\lesssim10^{-20}M_P.
	\label{decayX}
\end{equation}
As a consequence, the reheating temperature is estimated as $T_r\sim0.4\sqrt{\Gamma M_P}\lesssim10^8~\mathrm{GeV}$, which is far less than the inflation energy scale. 

With a specific reheating mechanism, the evolution of inflaton field obeys the following equation
\begin{equation}
	\Phi''+2\mathcal H\Phi'+a^2\frac{\partial V}{\partial\Phi}+a\Gamma\Phi'=0.
	\label{phieq}
\end{equation}
Then we can write down the energy density of the inflaton
\begin{equation}
	\rho_\Phi=\frac{\Phi'^2}{2a^2}+V(\Phi),
	\label{rhophi}
\end{equation}
and the evolving equation for $\psi$ and SM particles as radiation
\begin{equation}
	\rho_r'+4\mathcal{H}\rho_r-\frac{\Gamma\Phi'^2}{a}=0.
	\label{rhor}
\end{equation}
Note that the total energy density in the Friedmann equation is the sum of $\rho_\Phi$ and $\rho_r$. So combining Eq.~(\ref{Feq}) with (\ref{phieq})-(\ref{rhor}) and (\ref{V}), we can obtain a numerical solution of $a(\tau)$, which describes the evolution of cosmic scale within the period of inflation, reheating, and the ensuing radiation-dominated epoch. 

\subsubsection{Non-perturbative production}

We take $\Gamma(\varphi\rightarrow\psi\psi)=10^{-20}M_P$ as an example in this subsection, which is the maximum rate allowed in the toy model we constructed above. We shall keep in mind that generally the reheating temperature is free as long as it satisfies the bound from BBN and energy density limit after inflation. 

\begin{figure}
	\centering
	\includegraphics[width=0.75\textwidth]{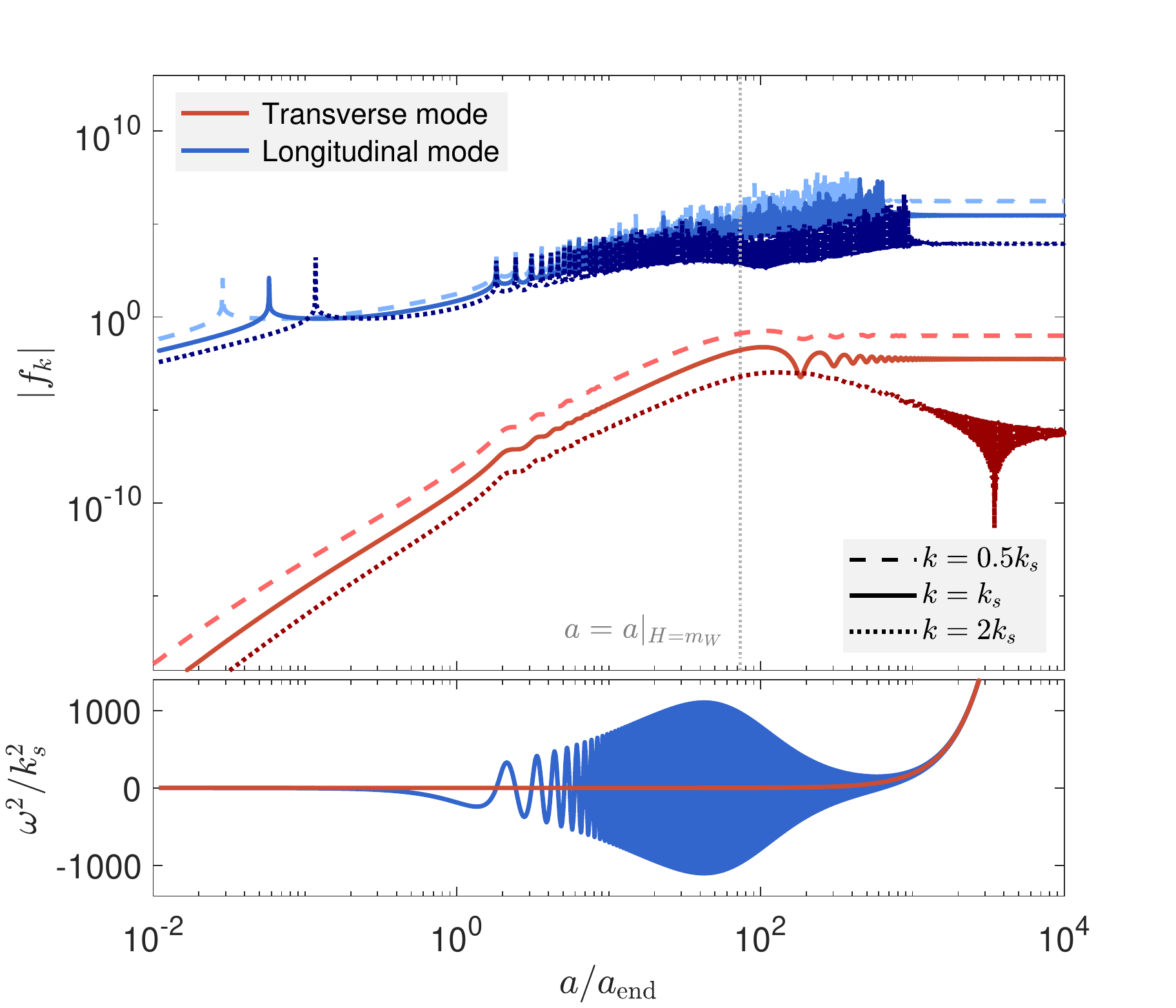}
	\caption{Evolution of $|f_k|$ and $\omega^2$ for the transverse mode (red line, solved by the adiabatic approximation) and longitudinal mode (blue line, solved directly in Eq.~(\ref{oEqWL})) with $m_W=10^{10}~\mathrm{GeV}$, $\Gamma=10^{-20}M_P$, and $k=0.5,1,2~k_s$ ($k_s$ denotes the dominate mode of production). It depicts an increasing abundance of dark matters in the co-moving volume until the Hubble parameter becomes lower than the mass of WGB. Moreover, the existence of tachyonic enhancement makes the creation of the longitudinal mode far more productive than the transverse mode.}
	\label{f_and_w2}
\end{figure}

We solve the numerical result of $a(\tau)$, and substitute it into Eq.~(\ref{oEqWT})-(\ref{w2WL}), the mode functions for each eventual mass $m_W$ (defined as Eq.~(\ref{mW}) or $\tilde m|_{\Phi=\Phi_0}$) and co-moving momentum $k$ can be derived with an initial condition. Because all the $k$-modes we are interested in once had the wavelength much smaller than the horizon scale in the far past of the inflation, it is then expected the term $k^2$ is dominant in Eq.~(\ref{w2WT}) and (\ref{w2WL}) under the limit of $\tau\rightarrow-\infty$. Therefore, the solution to the equation of harmonic oscillator is suggested to be an initial condition for both the transverse mode and the longitudinal mode, that is the well-known Bunch-Davies vacuum state \cite{BDvacuum}
\begin{equation}
	\lim_{\tau\rightarrow-\infty}W_\pm(\mathrm{or}~\mathcal W_L)=\frac{1}{\sqrt{2k}}e^{-ik\tau}.
	\label{inicon}
\end{equation}
With this initial condition, we illustrate with the eventual mass $m_W=10^{10}~\mathrm{GeV}$ in Fig.~\ref{f_and_w2}, where $f_k$ is defined as a function that is proportional to the energy density Eq.~(\ref{rhoT}) and (\ref{rhoL})
\begin{equation}
	\begin{aligned}
		f_{k,\pm}\equiv&\left[|W'_\pm|^2+(k^2+a^2\tilde m^2)|W_\pm|^2\right]/2\omega_\pm-\frac{1}{2}=|\mathcal B^k_{\pm}|^2,\\
		f_{k,L}\equiv&\bigg[|\mathcal W'_L|^2+(k^2+a^2\tilde m^2)|\mathcal W_L|^2+\bigg(\frac{(\frac{a'}{a}+\frac{\tilde m'}{\tilde m})k^2}{k^2+a^2\tilde m^2}\bigg)^2|\mathcal W_L|^2\\
		&-\frac{(\frac{a'}{a}+\frac{\tilde m'}{\tilde m})k^2}{k^2+a^2\tilde m^2}(\mathcal W'_L\mathcal W^*_L+\mathcal W'^*_L\mathcal W_L)\bigg]/2\omega_L-\frac{1}{2}=|\mathcal B^k_{L}|^2,
		\label{fk}
	\end{aligned}
\end{equation}
and $a_\mathrm{end}$ denotes the scale factor when the inflation ends (defined as the moment that the slow-roll parameter $\epsilon=1$). Moreover, $k_s$ denotes the dominate mode of production, which is approximately equal to the product of $m_W$ and the scale factor $a$ for the moment that the Hubble parameter $H=m_W$ (if $m_W>H_{\mathrm{inf}}$, the dominant mode is about $k\sim a_\mathrm{end}H_{\mathrm{inf}}$).

There are two interesting aspects in Fig.~\ref{f_and_w2}. Firstly, the production of the longitudinal mode is more abundantly than the transverse one, which has been shown by \cite{vdm1, pgdm2}. This behavior can be understood as follows. When $a\ll a_\mathrm{end}\lesssim k/m_W$, the frequency $\omega_\pm$ of the transverse mode is almost constant, which leads to a non-increasing mode function as plane wave. However, for the longitudinal mode, we can show that $\omega_L^2\sim k^2-a''/a$ with $a\ll a_\mathrm{end}\lesssim k/m_W$, here we ignore the impact of changing mass for convenience. Since the universe is approximately regarded as the de Sitter space during the inflationary epoch, that is $a\sim-1/H_{\mathrm{inf}}\tau$, we have $\omega_L^2\sim k^2-2a^2H_{\mathrm{inf}}^2$, which decreases as the universe expands. Substituting it into Eq.~(\ref{oEqWL}), then we obtain an increasing mode function
\begin{equation}
	\mathcal W_L|_{a\ll k/m_W}\sim\frac{e^{-ik\tau}}{\sqrt{2k}}\left(1+\frac{iaH_{\mathrm{inf}}}{k}\right).
	\label{WLinf}
\end{equation}
It indicates that when $a\ll k/H_{\mathrm{inf}}$, the mode function of the longitudinal mode is also a plane wave like the transverse one. But once the scale factor becomes large enough to make the negative $\omega_L^2$ emerge, the imaginary part of Eq.~(\ref{WLinf}) will surge exponentially. This situation is usually called the tachyonic enhancement, which leads to the high yield of the longitudinal mode compared to the transverse one.

The second noticeable aspect is that the function $f_k$ for the dominant mode $k_s$ gradually approaches a stationary value after the Hubble parameter $H$ becoming smaller than $m_W$. This is because the term $a^2m_W^2$ in Eq.~(\ref{w2WT}) and (\ref{w2WL}) turns into a leading term after that moment, which leads to a steady solution with $f_k\propto\rho a^3\sim\mathrm{constant}$ (see \cite{vdm2}). And the particles become non-relativistic due to the expansion of the universe. We can choose an arbitrary moment that satisfies $H\ll m_W$ to evaluate the spectrum of $f_k(k)$ around the dominant mode $k_s$, and calculate the integral (\ref{rhoT}), (\ref{rhoL}) or (\ref{rhoB}) to obtain the energy density $\rho$ of WGB at that moment. In practice, we consider a moment that is later than the end of reheating as the conserved quantity $\rho/s$ can be defined for the WGB, where $s=\frac{2\pi^2}{45}g_{*S}(T)T^3$ is the total entropy density for the reheated particles, and $g_{*S}(T)$ denotes the effective number of degrees of freedom in entropy with temperature $T$. This quantity measures the relic abundance of non-perturbative produced particles.

\begin{figure}
	\centering
	\includegraphics[width=0.75\textwidth]{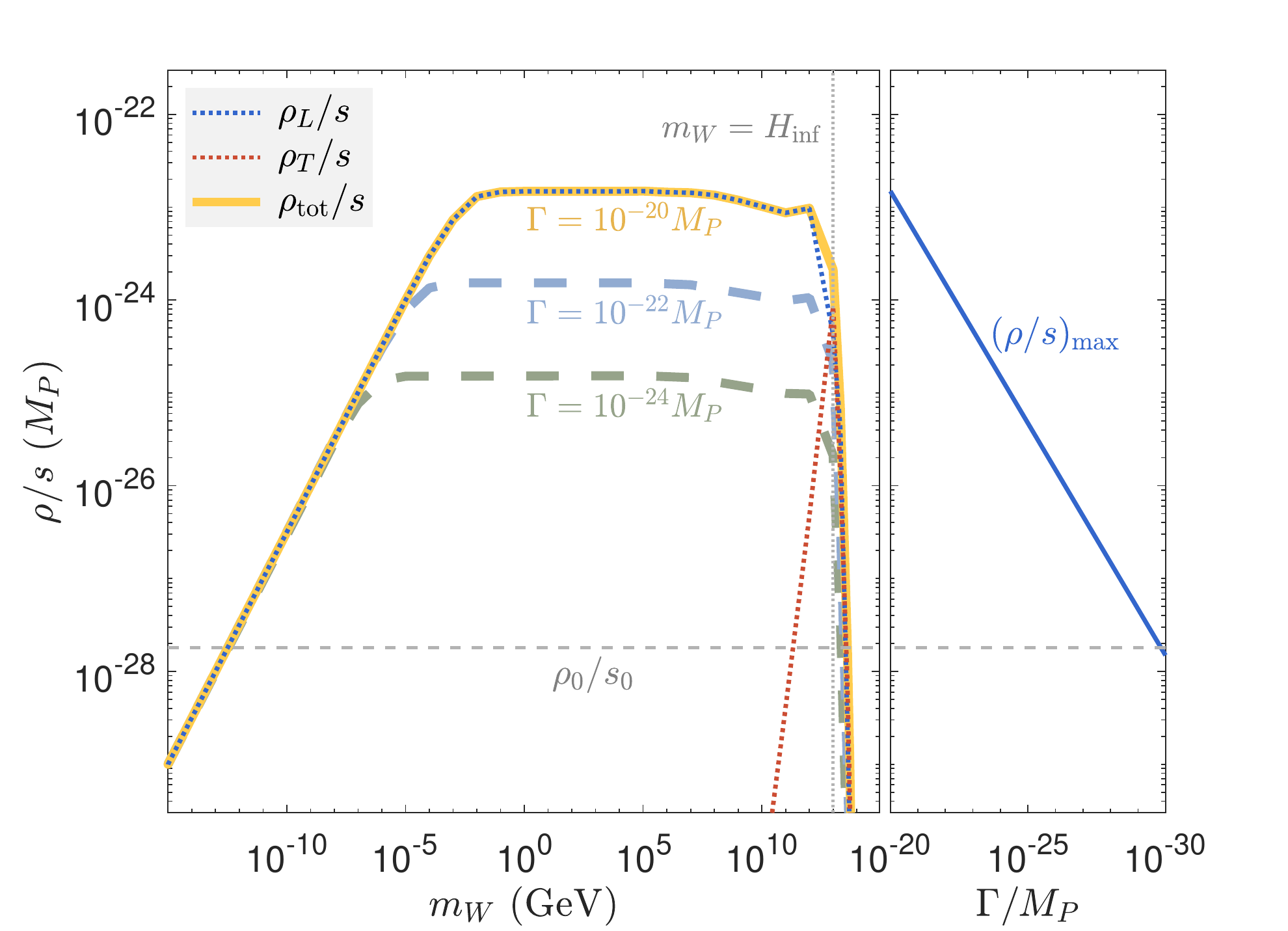}
	\caption{(Left) The relation between the relic abundance and the mass of non-perturbative produced WGB. The feasible mass of WGB should be restricted to the range that $\rho/s\leq\rho_0/s_0$. (Right) The maximal abundance decreases proportionately with the total decay rate of inflaton. It shows that if $\Gamma<10^{-30}M_P$ or $T_r<10^3~\mathrm{GeV}$, the non-perturbative produced WGB would only be part of dark matter.}
	\label{abundance}
\end{figure}

Fig.~\ref{abundance} shows how relic abundance varies with $m_W$ and decay constant $\Gamma$, where $m_W$ is determined by the parameters $\lambda$ and $g_W$, and the parameter $\beta$ has been fixed according to the constraint of observation on $H_{\mathrm{inf}}$ discussed in Sec.~II. We can see that there are three cases in the relationship between $\rho/s$ and $m_W$, which are summarized as
\begin{equation}
	\begin{aligned}
		\rho/s\propto
		\begin{cases}
			\sqrt{m_W},~&\mathrm{for}~m_W<H_{\mathrm{reh}},\\
			\mathrm{almost~constant},~&\mathrm{for}~H_{\mathrm{reh}}<m_W<H_{\mathrm{inf}},\\
			\mathrm{exponential~decrease},~&\mathrm{for}~m_W>H_{\mathrm{inf}},
		\end{cases}
		\label{ab_m}
	\end{aligned}
\end{equation}
where $H_{\mathrm{reh}}$ is the Hubble parameter at the end of reheating. 
For the current universe, $T_0=2.73~\mathrm{K}= 9.65\times10^{-32}M_P$, $g_{*S0}=3.91$, and the present energy density for the dark matters is $\rho_0=2.73\times10^{-121}M_P^4$, then we can derive $\rho_0/s_0=1.8\times10^{-28}M_P$. This observational result gives a constraint to the mass of WGB. It can be clearly identified in Fig.~\ref{abundance} that the relic abundance for the WGB produced in a non-perturbative pattern will be larger than that of dark matters in the current universe if the mass of WGB is in the interval of $3\times10^{-13}~\mathrm{GeV}$ to $\sim5\times10^{13}~\mathrm{GeV}$ (The upper limit is in the range of $10^{13}\mathrm{-}10^{14}~\mathrm{GeV}$, which is slightly dependent on the decay rate $\Gamma$). Therefore, the observation constrains two possible ranges of mass, $m_W\lesssim 3\times10^{-13}~\mathrm{GeV}$ or $m_W\gtrsim 5\times10^{13}~\mathrm{GeV}$, such that the relic density is not larger than the observation value.

When $m_W<3\times10^{-13}~\mathrm{GeV}$, according to the $\varphi-\tilde W$ interaction discussed in Sec.~II, there will be other production channels for WGB through the perturbative decay $\varphi\rightarrow\tilde W\tilde W$ and the annihilation $\varphi\varphi\rightarrow\tilde W\tilde W$. Since the contribution of decay is much larger than that of annihilation in our case, we only consider the decay. Based on Eq.~(\ref{m2expand}) and (\ref{mW}), the decay rate of $\varphi\rightarrow\tilde W\tilde W$ can be easily calculated as
\begin{equation}
	\begin{aligned}
		\Gamma(\varphi\rightarrow\tilde W\tilde W)&=\frac{3\sinh^2(2\Phi_0/\sqrt{6})g_W^4}{\pi}\frac{\sqrt{(m_\Phi/2)^2-m_W^2}}{m_\Phi^2}\left(3+\frac{m_\Phi^4}{4m_W^4}-\frac{m_\Phi^2}{m_W^2}\right)\\
		&\sim\frac{3m_\Phi^3}{288\pi}\frac{\sinh^2(2\Phi_0/\sqrt{6})}{\cosh^4(\Phi_0/\sqrt{6})},~\mathrm{when}~m_W\ll m_\Phi.
		\label{decayW}
	\end{aligned}
\end{equation}
Note that the mass of inflaton is around $m_\Phi\sim 2\times10^{13}~\mathrm{GeV}\sim10^{-5}M_P$ in this model, and the typical field value $\Phi_0\sim 10M_P$, thus the decay rate can be estimated as $\Gamma(\varphi\rightarrow\tilde W\tilde W)\sim10^{-17}M_P$. Comparing to Eq.~(\ref{decayX}), the decay rate of $\varphi\rightarrow\tilde W\tilde W$ is far larger than that of $\varphi\rightarrow\psi\psi$. This situation will lead to an inadequate reheating to SM, and result in additional dark radiation in our universe. In such a case, we would need other reheating mechanism that can transfer most of energy in inflaton field into other particles, and subsequently SM particles.

For the large mass range $m_W\gtrsim5\times10^{13}~\mathrm{GeV}$, there is a $-\psi^2 W^\mu W_\mu$ term due to the covariant derivative $D_\mu=\partial_\mu-W_\mu$ in Eq.~(\ref{reh}). This implies that the annihilation $\psi\psi\rightarrow \tilde W\tilde W$ is allowed in the large mass case. However, if the reheating temperature is too low, then the annihilation rate will be negligible, because the number density $n$ of $\psi$ with enough energy to participate in the production of WGB is exponentially reduced with decreasing temperature $T$, $n\propto\exp(-E/T)$. Therefore, with a low-temperature reheating mechanism, the non-perturbative production can reasonably provide enough abundance for WGB as dark matter with the mass $m_W\sim5\times10^{13}~\mathrm{GeV}$.

Another visible aspect in Fig.~\ref{abundance} is that the decay rate $\Gamma$ strongly affects the non-perturbative production. Note that $\Gamma$ denotes the total rate of all decay channels of inflaton, including decay to SM particles and to dark matters. Hence if $m_W<m_\Phi/2$, the $\varphi\rightarrow\tilde W\tilde W$ channel is open, and the total decay rate is never lower than $\Gamma(\varphi\rightarrow\tilde W\tilde W)$. Then if $m_W>m_\Phi/2$, the total decay rate is equivalent to $\Gamma(\varphi\rightarrow\psi\psi)$ as the reheating rate. It is apparent on the right of Fig.~\ref{abundance} that the maximal abundance of non-perturbative produced WGB decreases proportionately with $\Gamma$, and if $\Gamma<10^{-30}M_P$, the non-perturbative pattern is unable to produce sufficient WGB as dark matter. Since we have shown $\Gamma(\varphi\rightarrow\tilde W\tilde W)\sim10^{-17}M_P$ in this model (for $m_W<m_\Phi/2$), the $\Gamma<10^{-30}M_P$ case only appears when $m_W>m_\Phi/2$, which corresponds to the reheating temperature $T_r\sim0.4\sqrt{\Gamma M_P}<10^3~\mathrm{GeV}$. As a result, for this range of $T_r$, only heavy WGB can be a dark matter candidate, but with insufficient abundance.

\subsubsection{High-temperature reheating}

The preceding discussions consider a low-temperature reheating mechanism that is constructed for simplicity. In this section, we investigate the production of WGB with a high-temperature of reheating.


Let us focus on the small mass range first. The perturbative pattern of production is primarily the decay of inflaton $\varphi\rightarrow\tilde W\tilde W$. The annihilation $\psi\psi\rightarrow \tilde W\tilde W$ is also present ($\psi$ denotes an arbitrary scalar in the thermal equilibrium after reheating), but it is negligible, because the coupling of this process is too small according to the relation $g_W\propto m_W$ under our definition. Hence we only need to consider the decay. In previous sections, we have derived that the decay rate of $\varphi\rightarrow\tilde W\tilde W$ is $\Gamma_W\sim10^{-17}M_P$. If this decay rate is far larger than that to the SM particles, $\Gamma_W\gg\Gamma_{SM}$, the WGB can contribute as dark radiation. Therefore, the upper limit of abundance of dark radiation constrains the lower limit of $\Gamma_{SM}$ or $T_r$. Substituting $\Gamma=\Gamma_W+\Gamma_{SM}$ into Eq.~(\ref{phieq}), and combining it with the equation
\begin{equation}
	\rho_{SM/W}'+4\mathcal{H}\rho_{SM/W}-\frac{\Gamma_{SM/W}\Phi'^2}{a}=0,
	\label{rhoSMW}
\end{equation}
we can derive the energy density of dark radiation $\rho_W$ after reheating. Then the amount of dark radiation at present (denoted as $\Omega_{\mathrm{DR}}$) can be estimated. The result is, if $\Omega_{\mathrm{DR}}<\Omega_{\mathrm{SM}}$ is expected at matter-radiation equality, the magnitude of $\Gamma_{SM}$ cannot be lower than the level of $\Gamma_W$, which restricts the reheating temperature to be $T_r>4\times10^9~\mathrm{GeV}$. In conclusion, under this constraint, the small mass WGB can be the dark matter candidate, which embodies a tiny amount of dark radiation produced in the inflaton decay and the majority of cold component originates from the vacuum fluctuation during inflation. Meanwhile, if it contributes to the whole abundance of dark matters, the mass should be around $m_W\sim3\times10^{-13}~\mathrm{GeV}$.

For the heavy WGB case, the decay process is forbidden kinematically. However, the annihilation channel becomes significant, which is a kind of ``freeze-in'' mechanism of production \cite{freezein1, freezein2, FIMP}. There are two main channels for annihilation, direct annihilation owing to the $\psi^2W^\mu W_\mu$ coupling, and the other by gravitational interaction~\cite{Garny:2015sjg, Tang:2016vch, pgdm3, Garny:2017kha}. The amplitude $\mathcal A\equiv\sum|\mathcal M|^2$ of them can be calculated as follows
\begin{equation}
	\mathcal A(\psi\psi\rightarrow\tilde W\tilde W)=16g_W^4\left(3+\frac{\bar{\bm s}(\bar{\bm s}/4-m_W^2)}{m_W^4}\right),
	\label{A1}
\end{equation}
\begin{equation}
	\begin{aligned}
		\mathcal A(\psi\psi\overset{\mathrm{grav}}{\rightarrow}\tilde W\tilde W)=\kappa^4\left(\frac{101m_W^4m_\psi^4}{60\bar{\bm s}^2}-\frac{m_W^2m_\psi^2}{20\bar{\bm s}}(11m_\psi^2+m_W^2)\right.\\
		\left.+\frac{1}{240}(19m_\psi^4+76m_W^2m_\psi^2)-\frac{7\bar{\bm s}}{240}(m_W^2+m_\psi^2)+\frac{\bar{\bm s}^2}{160}\right),
		\label{A2}
	\end{aligned}
\end{equation}
where $\bar{\bm s}\equiv(p_{\psi1}+p_{\psi2})^2$, $\kappa\equiv\sqrt{32\pi G}=2$. Then the total cross section can be easily derived as
\begin{equation}
	\sigma_{\mathrm{tot}}=\frac{1}{16\pi\bar{\bm s}tg_i^2}\frac{|\vec p_f|}{|\vec p_i|}\mathcal A_{\mathrm{tot}},
	\label{CrossSection}
\end{equation}
where $t=2$ is the symmetric factor for the identical final states, $g_i$ is the degree of freedom for initial state. Finally, the relic abundance of dark matters produced in annihilation process can be calculated with the following formula (see also in \cite{pgdm3})
\begin{equation}
	Y=\int_{\sim0}^{\sim T_r}\frac{dT}{HTs}\left[\frac{Tg_i^2}{32\pi^4}\int d\bar{\bm s}\sigma_{\mathrm{tot}}\sqrt{\bar{\bm s}}(\bar{\bm s}-4m_\psi^2)K_1\left(\frac{\sqrt{\bar{\bm s}}}{T}\right)\right],
	\label{Y}
\end{equation}
where $H$ is the Hubble parameter, $s$ is the total entropy density mentioned earlier, and $K_1$ is the modified Bessel function of the second kind with order one. It is seen that for heavy particles, larger relic abundance of dark matter would result from higher reheating temperature. 

\begin{figure}
	\centering
	\includegraphics[width=0.75\textwidth]{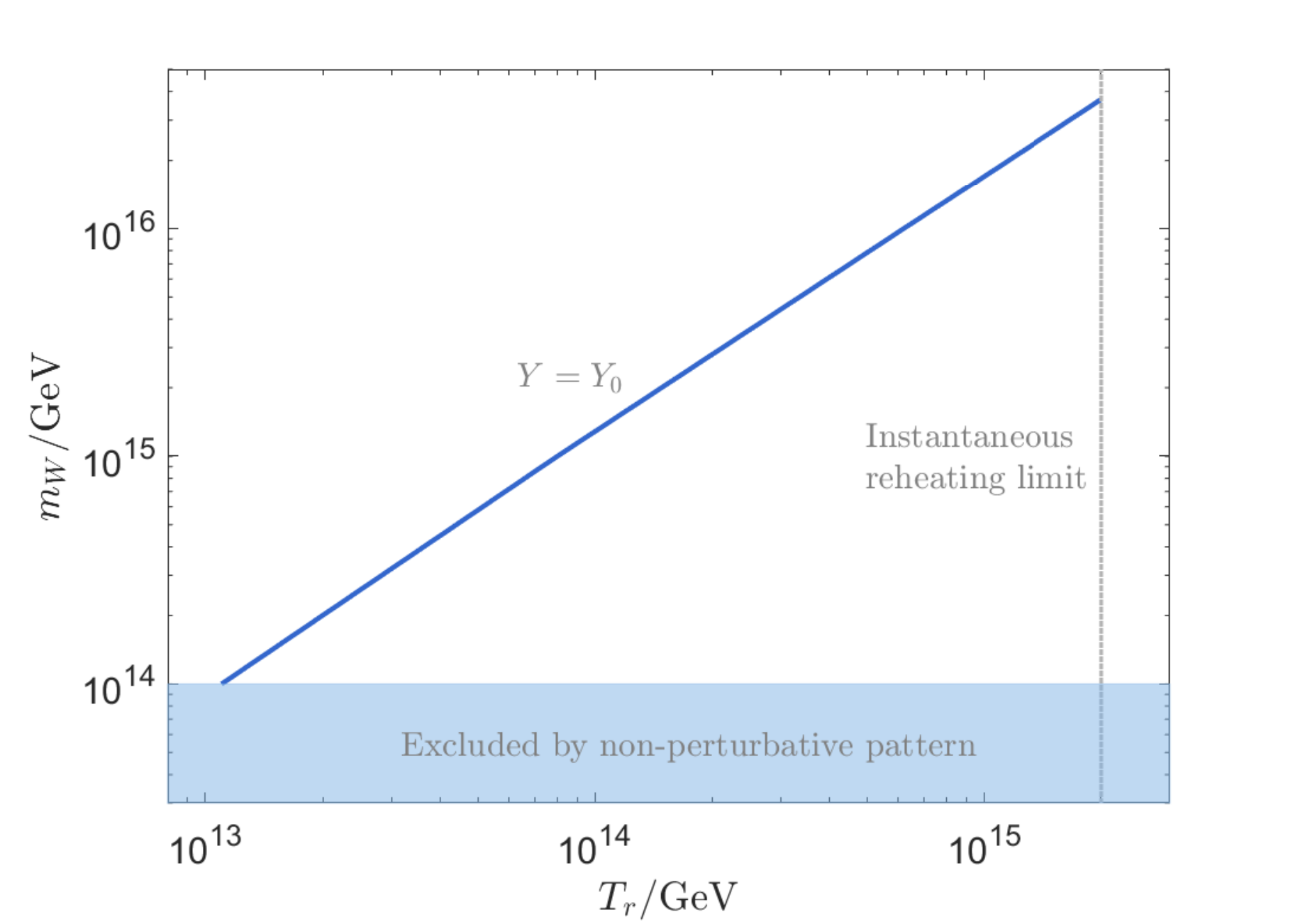}
	\caption{The proper mass $m_W$ for the ``freeze in'' production (annihilation channel) that derives an observation-consistent relic abundance is approximately linear with the reheating temperature $T_r$, where the $m_W<10^{14}~\mathrm{GeV}$ range is over-produced due to non-perturbative production. Under the instantaneous reheating limit, the WGB can possess an enormous mass of up to $4\times10^{16}~\mathrm{GeV}$.}
	\label{annihilation}
\end{figure}

With supposing $m_\psi\ll m_W$, we can calculate the relic abundance $Y$ for each set of $T_r$ and $m_W$. Then comparing it with present abundance of dark matter $Y_0\sim5\times10^{-9}/m_W$, we can get the relation between $m_W$ and $T_r$ for $Y=Y_0$, as shown in Fig.~\ref{annihilation}. It shows that the WGB in the large mass range is feasible to be the dark matter candidate with abundance that is consistent with observation as long as the reheating temperature $T_r>10^{13}~\mathrm{GeV}$, and the $m_W$ corresponding to the proper abundance is approximately linear with $T_r$. Moreover, since the instantaneous reheating limit is estimated as $T_{\mathrm{max}}\sim(3H_{\mathrm{inf}}^2/g_{*S}(T_r))^{1/4}\sim2\times10^{15}~\mathrm{GeV}$, the WGB can reach enormous mass of up to $4\times10^{16}~\mathrm{GeV}$.

Now we can summarize the main results we have demonstrated in the preceding discussions. First, if the reheating temperature is too small, $T_r<10^3~\mathrm{GeV}$, the WGB can only serve as part of dark matter with insufficient abundance. Because the non-perturbative production is inefficient for all mass range, and the perturbative contribution is also not productive or forbidden by the observations. Then, for $10^3~\mathrm{GeV}<T_r\lesssim4\times10^9~\mathrm{GeV}$, only the non-perturbative production can reasonably provide enough abundance for WGB as dark matter, and only the large mass is allowed, $m_W\sim10^{13}\mathrm{-}10^{14}~\mathrm{GeV}$. Moreover, for $4\times10^9~\mathrm{GeV}\lesssim T_r<10^{13}~\mathrm{GeV}$, there are two feasible values of mass for WGB, the small mass $m_W\sim3\times10^{-13}~\mathrm{GeV}$ and the large mass $m_W\sim10^{14}~\mathrm{GeV}$. The production channel is also mainly the non-perturbative pattern. Finally, for $10^{13}~\mathrm{GeV}<T_r<T_{\mathrm{max}}$, the small mass case is similar to the foregoing one, which is restricted to $m_W\sim3\times10^{-13}~\mathrm{GeV}$, and produced mainly in the non-perturbative channel. But the permissible large mass case is different from the above, because the perturbative annihilation becomes dominant under extremely high $T_r$, which enables WGB with mass from $10^{14}~\mathrm{GeV}$ to $4\times10^{16}~\mathrm{GeV}$ to be a dark matter candidate.

\section{Conclusions}

We work on a scaling invariant theory of gravity with a quadratic scalar curvature, namely the Weyl $R^2$ gravity. The model contains a viable inflationary scenario that is different with the conventional Starobinsky $R^2$ model. We calculate the spectral index and tensor-to-scalar ratio, and confront them with the latest cosmological observations. Our results indicate that this model's predictions agree with the observations and its differences from Starobinsky model can be tested with future CMB experiment.

The model also contains a stable gauge boson, Weyl gauge boson (WGB), which can serve as a dark matter candidate. The breaking of local scaling symmetry makes WGB massive in the inflationary period, which causes an inevitable non-perturbative production for WGB, namely emerging from the inflationary quantum fluctuations. Our investigations demonstrate that the relic abundance of non-perturbative produced WGB is related to the mass of WGB $m_W$ and the decay rate $\Gamma$ of inflaton. By comparing with the observation, we give the possible viable mass range of WGB, $m_W\sim 3\times10^{-13}~\mathrm{GeV}$ or $m_W>H_{\mathrm{inf}}$. Moreover, we also give the lower limit of decay width of the inflaton $\Gamma$ in this model, which corresponds to the reheating temperature $T_r>10^3~\mathrm{GeV}$.

We also demonstrate that there is a specific coupling between WGB and inflaton $\Phi$, which can lead to inflaton's decay into WGB if $m_W<m_\Phi/2$. This provides a stringent constraint on the allowable reheating temperature in the case of small $m_W$. Finally, for the case of extremely high reheating temperature $T_r$, the perturbative freeze-in channel becomes important and the dominant contribution, namely the annihilation of particles in the thermal bath into WGB through direct coupling or gravitational interaction. It opens new mass range of WGB up to $4\times10^{16}~\mathrm{GeV}$ as dark matter candidate.

\begin{acknowledgments}
YT is supported by National Key Research and Development Program of China (Grant No.2021YFC2201901), Natural Science Foundation of China (NSFC) under Grants No.~11851302, the Fundamental Research Funds for the Central Universities and Key Research Program of the Chinese Academy of Sciences, Grant No. XDPB15. YLW is supported in part by the National Key Research and Development Program of China under Grant No.2020YFC2201501, and NSFC under Grants No.~11851302, No.~11851303, No.~11690022, No.~11747601, and the Strategic Priority Research Program of the Chinese Academy of Sciences under Grant No. XDB23030100.	
\end{acknowledgments}



\begin{thebibliography}{83}%
	\makeatletter
	\providecommand \@ifxundefined [1]{%
		\@ifx{#1\undefined}
	}%
	\providecommand \@ifnum [1]{%
		\ifnum #1\expandafter \@firstoftwo
		\else \expandafter \@secondoftwo
		\fi
	}%
	\providecommand \@ifx [1]{%
		\ifx #1\expandafter \@firstoftwo
		\else \expandafter \@secondoftwo
		\fi
	}%
	\providecommand \natexlab [1]{#1}%
	\providecommand \enquote  [1]{``#1''}%
	\providecommand \bibnamefont  [1]{#1}%
	\providecommand \bibfnamefont [1]{#1}%
	\providecommand \citenamefont [1]{#1}%
	\providecommand \href@noop [0]{\@secondoftwo}%
	\providecommand \href [0]{\begingroup \@sanitize@url \@href}%
	\providecommand \@href[1]{\@@startlink{#1}\@@href}%
	\providecommand \@@href[1]{\endgroup#1\@@endlink}%
	\providecommand \@sanitize@url [0]{\catcode `\\12\catcode `\$12\catcode
		`\&12\catcode `\#12\catcode `\^12\catcode `\_12\catcode `\%12\relax}%
	\providecommand \@@startlink[1]{}%
	\providecommand \@@endlink[0]{}%
	\providecommand \url  [0]{\begingroup\@sanitize@url \@url }%
	\providecommand \@url [1]{\endgroup\@href {#1}{\urlprefix }}%
	\providecommand \urlprefix  [0]{URL }%
	\providecommand \Eprint [0]{\href }%
	\providecommand \doibase [0]{http://dx.doi.org/}%
	\providecommand \selectlanguage [0]{\@gobble}%
	\providecommand \bibinfo  [0]{\@secondoftwo}%
	\providecommand \bibfield  [0]{\@secondoftwo}%
	\providecommand \translation [1]{[#1]}%
	\providecommand \BibitemOpen [0]{}%
	\providecommand \bibitemStop [0]{}%
	\providecommand \bibitemNoStop [0]{.\EOS\space}%
	\providecommand \EOS [0]{\spacefactor3000\relax}%
	\providecommand \BibitemShut  [1]{\csname bibitem#1\endcsname}%
	\let\auto@bib@innerbib\@empty
	\bibitem [{\citenamefont {Rubin}\ and\ \citenamefont {Ford}(1970)}]{grc1}%
	\BibitemOpen
	\bibfield  {author} {\bibinfo {author} {\bibfnamefont {V.~C.}\ \bibnamefont
			{Rubin}}\ and\ \bibinfo {author} {\bibfnamefont {W.~K.}\ \bibnamefont {Ford},
			\bibfnamefont {Jr.}},\ }\href {\doibase 10.1086/150317} {\bibfield  {journal}
		{\bibinfo  {journal} {Astrophys. J.}\ }\textbf {\bibinfo {volume} {159}},\
		\bibinfo {pages} {379} (\bibinfo {year} {1970})}\BibitemShut {NoStop}%
	\bibitem [{\citenamefont {Allen}\ \emph {et~al.}(2003)\citenamefont {Allen},
		\citenamefont {Fabian}, \citenamefont {Schmidt},\ and\ \citenamefont
		{Ebeling}}]{lss1}%
	\BibitemOpen
	\bibfield  {author} {\bibinfo {author} {\bibfnamefont {S.~W.}\ \bibnamefont
			{Allen}}, \bibinfo {author} {\bibfnamefont {A.~C.}\ \bibnamefont {Fabian}},
		\bibinfo {author} {\bibfnamefont {R.~W.}\ \bibnamefont {Schmidt}}, \ and\
		\bibinfo {author} {\bibfnamefont {H.}~\bibnamefont {Ebeling}},\ }\href
	{\doibase 10.1046/j.1365-8711.2003.06550.x} {\bibfield  {journal} {\bibinfo
			{journal} {Mon. Not. Roy. Astron. Soc.}\ }\textbf {\bibinfo {volume} {342}},\
		\bibinfo {pages} {287} (\bibinfo {year} {2003})},\ \Eprint
	{http://arxiv.org/abs/astro-ph/0208394} {arXiv:astro-ph/0208394} \BibitemShut
	{NoStop}%
	\bibitem [{\citenamefont {Refregier}(2003)}]{lensing1}%
	\BibitemOpen
	\bibfield  {author} {\bibinfo {author} {\bibfnamefont {A.}~\bibnamefont
			{Refregier}},\ }\href {\doibase 10.1146/annurev.astro.41.111302.102207}
	{\bibfield  {journal} {\bibinfo  {journal} {Ann. Rev. Astron. Astrophys.}\
		}\textbf {\bibinfo {volume} {41}},\ \bibinfo {pages} {645} (\bibinfo {year}
		{2003})},\ \Eprint {http://arxiv.org/abs/astro-ph/0307212}
	{arXiv:astro-ph/0307212} \BibitemShut {NoStop}%
	\bibitem [{\citenamefont {Tyson}\ \emph {et~al.}(1998)\citenamefont {Tyson},
		\citenamefont {Kochanski},\ and\ \citenamefont {Dell'Antonio}}]{lensing2}%
	\BibitemOpen
	\bibfield  {author} {\bibinfo {author} {\bibfnamefont {J.~A.}\ \bibnamefont
			{Tyson}}, \bibinfo {author} {\bibfnamefont {G.~P.}\ \bibnamefont
			{Kochanski}}, \ and\ \bibinfo {author} {\bibfnamefont {I.~P.}\ \bibnamefont
			{Dell'Antonio}},\ }\href {\doibase 10.1086/311314} {\bibfield  {journal}
		{\bibinfo  {journal} {Astrophys. J. Lett.}\ }\textbf {\bibinfo {volume}
			{498}},\ \bibinfo {pages} {L107} (\bibinfo {year} {1998})},\ \Eprint
	{http://arxiv.org/abs/astro-ph/9801193} {arXiv:astro-ph/9801193} \BibitemShut
	{NoStop}%
	\bibitem [{\citenamefont {Fields}\ \emph {et~al.}(2020)\citenamefont {Fields},
		\citenamefont {Molaro},\ and\ \citenamefont {Sarkar}}]{nucl}%
	\BibitemOpen
	\bibfield  {author} {\bibinfo {author} {\bibfnamefont {B.}~\bibnamefont
			{Fields}}, \bibinfo {author} {\bibfnamefont {P.}~\bibnamefont {Molaro}}, \
		and\ \bibinfo {author} {\bibfnamefont {S.}~\bibnamefont {Sarkar}} (\bibinfo
		{collaboration} {Particle Data Group Collaboration}),\ }\href@noop {}
	{\bibfield  {journal} {\bibinfo  {journal} {Prog. Theor. Exp. Phys.}\
		}\textbf {\bibinfo {volume} {2020}},\ \bibinfo {pages} {083C01} (\bibinfo
		{year} {2020})}\BibitemShut {NoStop}%
	\bibitem [{\citenamefont {Aghanim}\ \emph {et~al.}(2020)\citenamefont {Aghanim}
		\emph {et~al.}}]{Planck2018}%
	\BibitemOpen
	\bibfield  {author} {\bibinfo {author} {\bibfnamefont {N.}~\bibnamefont
			{Aghanim}} \emph {et~al.} (\bibinfo {collaboration} {Planck}),\ }\href
	{\doibase 10.1051/0004-6361/201833910} {\bibfield  {journal} {\bibinfo
			{journal} {Astron. Astrophys.}\ }\textbf {\bibinfo {volume} {641}},\ \bibinfo
		{pages} {A6} (\bibinfo {year} {2020})},\ \bibinfo {note} {[Erratum:
		Astron.Astrophys. 652, C4 (2021)]},\ \Eprint
	{http://arxiv.org/abs/1807.06209} {arXiv:1807.06209 [astro-ph.CO]}
	\BibitemShut {NoStop}%
	\bibitem [{\citenamefont {Lin}(2019)}]{detect1}%
	\BibitemOpen
	\bibfield  {author} {\bibinfo {author} {\bibfnamefont {T.}~\bibnamefont
			{Lin}},\ }\href {\doibase 10.22323/1.333.0009} {\bibfield  {journal}
		{\bibinfo  {journal} {PoS}\ }\textbf {\bibinfo {volume} {333}},\ \bibinfo
		{pages} {009} (\bibinfo {year} {2019})},\ \Eprint
	{http://arxiv.org/abs/1904.07915} {arXiv:1904.07915 [hep-ph]} \BibitemShut
	{NoStop}%
	\bibitem [{\citenamefont {Schumann}(2019)}]{detect2}%
	\BibitemOpen
	\bibfield  {author} {\bibinfo {author} {\bibfnamefont {M.}~\bibnamefont
			{Schumann}},\ }\href {\doibase 10.1088/1361-6471/ab2ea5} {\bibfield
		{journal} {\bibinfo  {journal} {J. Phys. G}\ }\textbf {\bibinfo {volume}
			{46}},\ \bibinfo {pages} {103003} (\bibinfo {year} {2019})},\ \Eprint
	{http://arxiv.org/abs/1903.03026} {arXiv:1903.03026 [astro-ph.CO]}
	\BibitemShut {NoStop}%
	\bibitem [{\citenamefont {Billard}\ \emph {et~al.}(2021)\citenamefont {Billard}
		\emph {et~al.}}]{detect3}%
	\BibitemOpen
	\bibfield  {author} {\bibinfo {author} {\bibfnamefont {J.}~\bibnamefont
			{Billard}} \emph {et~al.},\ }\href@noop {} {\  (\bibinfo {year} {2021})},\
	\Eprint {http://arxiv.org/abs/2104.07634} {arXiv:2104.07634 [hep-ex]}
	\BibitemShut {NoStop}%
	\bibitem [{\citenamefont {Tang}\ and\ \citenamefont
		{Wu}(2020{\natexlab{a}})}]{WGB1}%
	\BibitemOpen
	\bibfield  {author} {\bibinfo {author} {\bibfnamefont {Y.}~\bibnamefont
			{Tang}}\ and\ \bibinfo {author} {\bibfnamefont {Y.-L.}\ \bibnamefont {Wu}},\
	}\href {\doibase 10.1016/j.physletb.2020.135320} {\bibfield  {journal}
		{\bibinfo  {journal} {Phys. Lett. B}\ }\textbf {\bibinfo {volume} {803}},\
		\bibinfo {pages} {135320} (\bibinfo {year} {2020}{\natexlab{a}})},\ \Eprint
	{http://arxiv.org/abs/1904.04493} {arXiv:1904.04493 [hep-ph]} \BibitemShut
	{NoStop}%
	\bibitem [{\citenamefont {Tang}\ and\ \citenamefont
		{Wu}(2020{\natexlab{b}})}]{WGB2}%
	\BibitemOpen
	\bibfield  {author} {\bibinfo {author} {\bibfnamefont {Y.}~\bibnamefont
			{Tang}}\ and\ \bibinfo {author} {\bibfnamefont {Y.-L.}\ \bibnamefont {Wu}},\
	}\href {\doibase 10.1088/1475-7516/2020/03/067} {\bibfield  {journal}
		{\bibinfo  {journal} {JCAP}\ }\textbf {\bibinfo {volume} {03}},\ \bibinfo
		{pages} {067} (\bibinfo {year} {2020}{\natexlab{b}})},\ \Eprint
	{http://arxiv.org/abs/1912.07610} {arXiv:1912.07610 [hep-ph]} \BibitemShut
	{NoStop}%
	\bibitem [{\citenamefont {Wu}(2016)}]{Wu:2015wwa}%
	\BibitemOpen
	\bibfield  {author} {\bibinfo {author} {\bibfnamefont {Y.-L.}\ \bibnamefont
			{Wu}},\ }\href {\doibase 10.1103/PhysRevD.93.024012} {\bibfield  {journal}
		{\bibinfo  {journal} {Phys. Rev.}\ }\textbf {\bibinfo {volume} {D93}},\
		\bibinfo {pages} {024012} (\bibinfo {year} {2016})},\ \Eprint
	{http://arxiv.org/abs/1506.01807} {arXiv:1506.01807 [hep-th]} \BibitemShut
	{NoStop}%
	\bibitem [{\citenamefont {Wu}(2018)}]{Wu:2017urh}%
	\BibitemOpen
	\bibfield  {author} {\bibinfo {author} {\bibfnamefont {Y.-L.}\ \bibnamefont
			{Wu}},\ }\href {\doibase 10.1140/epjc/s10052-017-5504-3} {\bibfield
		{journal} {\bibinfo  {journal} {Eur. Phys. J.}\ }\textbf {\bibinfo {volume}
			{C78}},\ \bibinfo {pages} {28} (\bibinfo {year} {2018})},\ \Eprint
	{http://arxiv.org/abs/1712.04537} {arXiv:1712.04537 [hep-th]} \BibitemShut
	{NoStop}%
	\bibitem [{\citenamefont {Weyl}(1929)}]{Weyl1}%
	\BibitemOpen
	\bibfield  {author} {\bibinfo {author} {\bibfnamefont {H.}~\bibnamefont
			{Weyl}},\ }\href {\doibase 10.1007/BF01339504} {\bibfield  {journal}
		{\bibinfo  {journal} {Z. Phys.}\ }\textbf {\bibinfo {volume} {56}},\ \bibinfo
		{pages} {330} (\bibinfo {year} {1929})}\BibitemShut {NoStop}%
	\bibitem [{\citenamefont {Mannheim}\ and\ \citenamefont
		{Kazanas}(1989)}]{Weyl3}%
	\BibitemOpen
	\bibfield  {author} {\bibinfo {author} {\bibfnamefont {P.~D.}\ \bibnamefont
			{Mannheim}}\ and\ \bibinfo {author} {\bibfnamefont {D.}~\bibnamefont
			{Kazanas}},\ }\href {\doibase 10.1086/167623} {\bibfield  {journal} {\bibinfo
			{journal} {Astrophys. J.}\ }\textbf {\bibinfo {volume} {342}},\ \bibinfo
		{pages} {635} (\bibinfo {year} {1989})}\BibitemShut {NoStop}%
	\bibitem [{\citenamefont {Romero}\ \emph {et~al.}(2012)\citenamefont {Romero},
		\citenamefont {Fonseca-Neto},\ and\ \citenamefont {Pucheu}}]{Weyl4}%
	\BibitemOpen
	\bibfield  {author} {\bibinfo {author} {\bibfnamefont {C.}~\bibnamefont
			{Romero}}, \bibinfo {author} {\bibfnamefont {J.~B.}\ \bibnamefont
			{Fonseca-Neto}}, \ and\ \bibinfo {author} {\bibfnamefont {M.~L.}\
			\bibnamefont {Pucheu}},\ }\href {\doibase 10.1088/0264-9381/29/15/155015}
	{\bibfield  {journal} {\bibinfo  {journal} {Class. Quant. Grav.}\ }\textbf
		{\bibinfo {volume} {29}},\ \bibinfo {pages} {155015} (\bibinfo {year}
		{2012})},\ \Eprint {http://arxiv.org/abs/1201.1469} {arXiv:1201.1469 [gr-qc]}
	\BibitemShut {NoStop}%
	\bibitem [{\citenamefont {Bars}\ \emph {et~al.}(2014)\citenamefont {Bars},
		\citenamefont {Steinhardt},\ and\ \citenamefont {Turok}}]{Weyl5}%
	\BibitemOpen
	\bibfield  {author} {\bibinfo {author} {\bibfnamefont {I.}~\bibnamefont
			{Bars}}, \bibinfo {author} {\bibfnamefont {P.}~\bibnamefont {Steinhardt}}, \
		and\ \bibinfo {author} {\bibfnamefont {N.}~\bibnamefont {Turok}},\ }\href
	{\doibase 10.1103/PhysRevD.89.043515} {\bibfield  {journal} {\bibinfo
			{journal} {Phys. Rev. D}\ }\textbf {\bibinfo {volume} {89}},\ \bibinfo
		{pages} {043515} (\bibinfo {year} {2014})},\ \Eprint
	{http://arxiv.org/abs/1307.1848} {arXiv:1307.1848 [hep-th]} \BibitemShut
	{NoStop}%
	\bibitem [{\citenamefont {Scholz}(2018)}]{Weyl6}%
	\BibitemOpen
	\bibfield  {author} {\bibinfo {author} {\bibfnamefont {E.}~\bibnamefont
			{Scholz}},\ }\href {\doibase 10.1007/978-1-4939-7708-6_11} {\bibfield
		{journal} {\bibinfo  {journal} {Einstein Stud.}\ }\textbf {\bibinfo {volume}
			{14}},\ \bibinfo {pages} {261} (\bibinfo {year} {2018})},\ \Eprint
	{http://arxiv.org/abs/1703.03187} {arXiv:1703.03187 [math.HO]} \BibitemShut
	{NoStop}%
	\bibitem [{\citenamefont {Ghilencea}(2019)}]{R2Weyl2}%
	\BibitemOpen
	\bibfield  {author} {\bibinfo {author} {\bibfnamefont {D.~M.}\ \bibnamefont
			{Ghilencea}},\ }\href {\doibase 10.1007/JHEP10(2019)209} {\bibfield
		{journal} {\bibinfo  {journal} {JHEP}\ }\textbf {\bibinfo {volume} {10}},\
		\bibinfo {pages} {209} (\bibinfo {year} {2019})},\ \Eprint
	{http://arxiv.org/abs/1906.11572} {arXiv:1906.11572 [gr-qc]} \BibitemShut
	{NoStop}%
	\bibitem [{\citenamefont {Ferreira}\ \emph {et~al.}(2019)\citenamefont
		{Ferreira}, \citenamefont {Hill}, \citenamefont {Noller},\ and\ \citenamefont
		{Ross}}]{R2Weyl3}%
	\BibitemOpen
	\bibfield  {author} {\bibinfo {author} {\bibfnamefont {P.~G.}\ \bibnamefont
			{Ferreira}}, \bibinfo {author} {\bibfnamefont {C.~T.}\ \bibnamefont {Hill}},
		\bibinfo {author} {\bibfnamefont {J.}~\bibnamefont {Noller}}, \ and\ \bibinfo
		{author} {\bibfnamefont {G.~G.}\ \bibnamefont {Ross}},\ }\href {\doibase
		10.1103/PhysRevD.100.123516} {\bibfield  {journal} {\bibinfo  {journal}
			{Phys. Rev. D}\ }\textbf {\bibinfo {volume} {100}},\ \bibinfo {pages}
		{123516} (\bibinfo {year} {2019})},\ \Eprint
	{http://arxiv.org/abs/1906.03415} {arXiv:1906.03415 [gr-qc]} \BibitemShut
	{NoStop}%
	\bibitem [{\citenamefont {Tang}\ and\ \citenamefont
		{Wu}(2020{\natexlab{c}})}]{R2Weyl4}%
	\BibitemOpen
	\bibfield  {author} {\bibinfo {author} {\bibfnamefont {Y.}~\bibnamefont
			{Tang}}\ and\ \bibinfo {author} {\bibfnamefont {Y.-L.}\ \bibnamefont {Wu}},\
	}\href {\doibase 10.1016/j.physletb.2020.135716} {\bibfield  {journal}
		{\bibinfo  {journal} {Phys. Lett. B}\ }\textbf {\bibinfo {volume} {809}},\
		\bibinfo {pages} {135716} (\bibinfo {year} {2020}{\natexlab{c}})},\ \Eprint
	{http://arxiv.org/abs/2006.02811} {arXiv:2006.02811 [hep-ph]} \BibitemShut
	{NoStop}%
	\bibitem [{\citenamefont {Beltran~Jimenez}\ \emph {et~al.}(2016)\citenamefont
		{Beltran~Jimenez}, \citenamefont {Heisenberg},\ and\ \citenamefont
		{Koivisto}}]{R2Weyl1}%
	\BibitemOpen
	\bibfield  {author} {\bibinfo {author} {\bibfnamefont {J.}~\bibnamefont
			{Beltran~Jimenez}}, \bibinfo {author} {\bibfnamefont {L.}~\bibnamefont
			{Heisenberg}}, \ and\ \bibinfo {author} {\bibfnamefont {T.~S.}\ \bibnamefont
			{Koivisto}},\ }\href {\doibase 10.1088/1475-7516/2016/04/046} {\bibfield
		{journal} {\bibinfo  {journal} {JCAP}\ }\textbf {\bibinfo {volume} {04}},\
		\bibinfo {pages} {046} (\bibinfo {year} {2016})},\ \Eprint
	{http://arxiv.org/abs/1602.07287} {arXiv:1602.07287 [hep-th]} \BibitemShut
	{NoStop}%
	\bibitem [{\citenamefont {Oda}(2020)}]{R2Weyl5}%
	\BibitemOpen
	\bibfield  {author} {\bibinfo {author} {\bibfnamefont {I.}~\bibnamefont
			{Oda}},\ }\href {\doibase 10.1142/S0217732320503046} {\bibfield  {journal}
		{\bibinfo  {journal} {Mod. Phys. Lett. A}\ }\textbf {\bibinfo {volume}
			{35}},\ \bibinfo {pages} {2050304} (\bibinfo {year} {2020})},\ \Eprint
	{http://arxiv.org/abs/2006.10867} {arXiv:2006.10867 [hep-th]} \BibitemShut
	{NoStop}%
	\bibitem [{\citenamefont {Ghilencea}(2021)}]{R2Weyl6}%
	\BibitemOpen
	\bibfield  {author} {\bibinfo {author} {\bibfnamefont {D.~M.}\ \bibnamefont
			{Ghilencea}},\ }\href {\doibase 10.1140/epjc/s10052-021-09226-1} {\bibfield
		{journal} {\bibinfo  {journal} {Eur. Phys. J. C}\ }\textbf {\bibinfo {volume}
			{81}},\ \bibinfo {pages} {510} (\bibinfo {year} {2021})},\ \Eprint
	{http://arxiv.org/abs/2007.14733} {arXiv:2007.14733 [hep-th]} \BibitemShut
	{NoStop}%
	\bibitem [{\citenamefont {Cai}\ \emph {et~al.}(2021)\citenamefont {Cai},
		\citenamefont {Hao},\ and\ \citenamefont {Wang}}]{R2Weyl7}%
	\BibitemOpen
	\bibfield  {author} {\bibinfo {author} {\bibfnamefont {R.-G.}\ \bibnamefont
			{Cai}}, \bibinfo {author} {\bibfnamefont {Y.-S.}\ \bibnamefont {Hao}}, \ and\
		\bibinfo {author} {\bibfnamefont {S.-J.}\ \bibnamefont {Wang}},\ }\href@noop
	{} {\  (\bibinfo {year} {2021})},\ \Eprint {http://arxiv.org/abs/2110.14718}
	{arXiv:2110.14718 [gr-qc]} \BibitemShut {NoStop}%
	\bibitem [{\citenamefont {Parker}(1969)}]{Parker1}%
	\BibitemOpen
	\bibfield  {author} {\bibinfo {author} {\bibfnamefont {L.}~\bibnamefont
			{Parker}},\ }\href {\doibase 10.1103/PhysRev.183.1057} {\bibfield  {journal}
		{\bibinfo  {journal} {Phys. Rev.}\ }\textbf {\bibinfo {volume} {183}},\
		\bibinfo {pages} {1057} (\bibinfo {year} {1969})}\BibitemShut {NoStop}%
	\bibitem [{\citenamefont {Parker}(1971)}]{Parker2}%
	\BibitemOpen
	\bibfield  {author} {\bibinfo {author} {\bibfnamefont {L.}~\bibnamefont
			{Parker}},\ }\href {\doibase 10.1103/PhysRevD.3.346} {\bibfield  {journal}
		{\bibinfo  {journal} {Phys. Rev. D}\ }\textbf {\bibinfo {volume} {3}},\
		\bibinfo {pages} {346} (\bibinfo {year} {1971})},\ \bibinfo {note} {[Erratum:
		Phys.Rev.D 3, 2546--2546 (1971)]}\BibitemShut {NoStop}%
	\bibitem [{\citenamefont {Ford}(1987)}]{Ford}%
	\BibitemOpen
	\bibfield  {author} {\bibinfo {author} {\bibfnamefont {L.~H.}\ \bibnamefont
			{Ford}},\ }\href {\doibase 10.1103/PhysRevD.35.2955} {\bibfield  {journal}
		{\bibinfo  {journal} {Phys. Rev. D}\ }\textbf {\bibinfo {volume} {35}},\
		\bibinfo {pages} {2955} (\bibinfo {year} {1987})}\BibitemShut {NoStop}%
	\bibitem [{\citenamefont {Lyth}\ and\ \citenamefont {Roberts}(1998)}]{Lyth}%
	\BibitemOpen
	\bibfield  {author} {\bibinfo {author} {\bibfnamefont {D.~H.}\ \bibnamefont
			{Lyth}}\ and\ \bibinfo {author} {\bibfnamefont {D.}~\bibnamefont {Roberts}},\
	}\href {\doibase 10.1103/PhysRevD.57.7120} {\bibfield  {journal} {\bibinfo
			{journal} {Phys. Rev. D}\ }\textbf {\bibinfo {volume} {57}},\ \bibinfo
		{pages} {7120} (\bibinfo {year} {1998})},\ \Eprint
	{http://arxiv.org/abs/hep-ph/9609441} {arXiv:hep-ph/9609441} \BibitemShut
	{NoStop}%
	\bibitem [{\citenamefont {Chung}\ \emph {et~al.}(2001)\citenamefont {Chung},
		\citenamefont {Crotty}, \citenamefont {Kolb},\ and\ \citenamefont
		{Riotto}}]{gpdm}%
	\BibitemOpen
	\bibfield  {author} {\bibinfo {author} {\bibfnamefont {D.~J.~H.}\
			\bibnamefont {Chung}}, \bibinfo {author} {\bibfnamefont {P.}~\bibnamefont
			{Crotty}}, \bibinfo {author} {\bibfnamefont {E.~W.}\ \bibnamefont {Kolb}}, \
		and\ \bibinfo {author} {\bibfnamefont {A.}~\bibnamefont {Riotto}},\ }\href
	{\doibase 10.1103/PhysRevD.64.043503} {\bibfield  {journal} {\bibinfo
			{journal} {Phys. Rev. D}\ }\textbf {\bibinfo {volume} {64}},\ \bibinfo
		{pages} {043503} (\bibinfo {year} {2001})},\ \Eprint
	{http://arxiv.org/abs/hep-ph/0104100} {arXiv:hep-ph/0104100} \BibitemShut
	{NoStop}%
	\bibitem [{\citenamefont {Baer}\ \emph {et~al.}(2015)\citenamefont {Baer},
		\citenamefont {Choi}, \citenamefont {Kim},\ and\ \citenamefont
		{Roszkowski}}]{nonthermal}%
	\BibitemOpen
	\bibfield  {author} {\bibinfo {author} {\bibfnamefont {H.}~\bibnamefont
			{Baer}}, \bibinfo {author} {\bibfnamefont {K.-Y.}\ \bibnamefont {Choi}},
		\bibinfo {author} {\bibfnamefont {J.~E.}\ \bibnamefont {Kim}}, \ and\
		\bibinfo {author} {\bibfnamefont {L.}~\bibnamefont {Roszkowski}},\ }\href
	{\doibase 10.1016/j.physrep.2014.10.002} {\bibfield  {journal} {\bibinfo
			{journal} {Phys. Rept.}\ }\textbf {\bibinfo {volume} {555}},\ \bibinfo
		{pages} {1} (\bibinfo {year} {2015})},\ \Eprint
	{http://arxiv.org/abs/1407.0017} {arXiv:1407.0017 [hep-ph]} \BibitemShut
	{NoStop}%
	\bibitem [{\citenamefont {Graham}\ \emph {et~al.}(2016)\citenamefont {Graham},
		\citenamefont {Mardon},\ and\ \citenamefont {Rajendran}}]{vdm1}%
	\BibitemOpen
	\bibfield  {author} {\bibinfo {author} {\bibfnamefont {P.~W.}\ \bibnamefont
			{Graham}}, \bibinfo {author} {\bibfnamefont {J.}~\bibnamefont {Mardon}}, \
		and\ \bibinfo {author} {\bibfnamefont {S.}~\bibnamefont {Rajendran}},\ }\href
	{\doibase 10.1103/PhysRevD.93.103520} {\bibfield  {journal} {\bibinfo
			{journal} {Phys. Rev. D}\ }\textbf {\bibinfo {volume} {93}},\ \bibinfo
		{pages} {103520} (\bibinfo {year} {2016})},\ \Eprint
	{http://arxiv.org/abs/1504.02102} {arXiv:1504.02102 [hep-ph]} \BibitemShut
	{NoStop}%
	\bibitem [{\citenamefont {Ema}\ \emph {et~al.}(2018)\citenamefont {Ema},
		\citenamefont {Nakayama},\ and\ \citenamefont {Tang}}]{pgdm1}%
	\BibitemOpen
	\bibfield  {author} {\bibinfo {author} {\bibfnamefont {Y.}~\bibnamefont
			{Ema}}, \bibinfo {author} {\bibfnamefont {K.}~\bibnamefont {Nakayama}}, \
		and\ \bibinfo {author} {\bibfnamefont {Y.}~\bibnamefont {Tang}},\ }\href
	{\doibase 10.1007/JHEP09(2018)135} {\bibfield  {journal} {\bibinfo  {journal}
			{JHEP}\ }\textbf {\bibinfo {volume} {09}},\ \bibinfo {pages} {135} (\bibinfo
		{year} {2018})},\ \Eprint {http://arxiv.org/abs/1804.07471} {arXiv:1804.07471
		[hep-ph]} \BibitemShut {NoStop}%
	\bibitem [{\citenamefont {Ema}\ \emph {et~al.}(2019)\citenamefont {Ema},
		\citenamefont {Nakayama},\ and\ \citenamefont {Tang}}]{pgdm2}%
	\BibitemOpen
	\bibfield  {author} {\bibinfo {author} {\bibfnamefont {Y.}~\bibnamefont
			{Ema}}, \bibinfo {author} {\bibfnamefont {K.}~\bibnamefont {Nakayama}}, \
		and\ \bibinfo {author} {\bibfnamefont {Y.}~\bibnamefont {Tang}},\ }\href
	{\doibase 10.1007/JHEP07(2019)060} {\bibfield  {journal} {\bibinfo  {journal}
			{JHEP}\ }\textbf {\bibinfo {volume} {07}},\ \bibinfo {pages} {060} (\bibinfo
		{year} {2019})},\ \Eprint {http://arxiv.org/abs/1903.10973} {arXiv:1903.10973
		[hep-ph]} \BibitemShut {NoStop}%
	\bibitem [{\citenamefont {Ahmed}\ \emph {et~al.}(2020)\citenamefont {Ahmed},
		\citenamefont {Grzadkowski},\ and\ \citenamefont {Socha}}]{vdm2}%
	\BibitemOpen
	\bibfield  {author} {\bibinfo {author} {\bibfnamefont {A.}~\bibnamefont
			{Ahmed}}, \bibinfo {author} {\bibfnamefont {B.}~\bibnamefont {Grzadkowski}},
		\ and\ \bibinfo {author} {\bibfnamefont {A.}~\bibnamefont {Socha}},\ }\href
	{\doibase 10.1007/JHEP08(2020)059} {\bibfield  {journal} {\bibinfo  {journal}
			{JHEP}\ }\textbf {\bibinfo {volume} {08}},\ \bibinfo {pages} {059} (\bibinfo
		{year} {2020})},\ \Eprint {http://arxiv.org/abs/2005.01766} {arXiv:2005.01766
		[hep-ph]} \BibitemShut {NoStop}%
	\bibitem [{\citenamefont {Chung}\ \emph {et~al.}(2019)\citenamefont {Chung},
		\citenamefont {Kolb},\ and\ \citenamefont {Long}}]{Chung:2018ayg}%
	\BibitemOpen
	\bibfield  {author} {\bibinfo {author} {\bibfnamefont {D.~J.~H.}\
			\bibnamefont {Chung}}, \bibinfo {author} {\bibfnamefont {E.~W.}\ \bibnamefont
			{Kolb}}, \ and\ \bibinfo {author} {\bibfnamefont {A.~J.}\ \bibnamefont
			{Long}},\ }\href {\doibase 10.1007/JHEP01(2019)189} {\bibfield  {journal}
		{\bibinfo  {journal} {JHEP}\ }\textbf {\bibinfo {volume} {01}},\ \bibinfo
		{pages} {189} (\bibinfo {year} {2019})},\ \Eprint
	{http://arxiv.org/abs/1812.00211} {arXiv:1812.00211 [hep-ph]} \BibitemShut
	{NoStop}%
	\bibitem [{\citenamefont {Kolb}\ and\ \citenamefont
		{Long}(2021)}]{Kolb:2020fwh}%
	\BibitemOpen
	\bibfield  {author} {\bibinfo {author} {\bibfnamefont {E.~W.}\ \bibnamefont
			{Kolb}}\ and\ \bibinfo {author} {\bibfnamefont {A.~J.}\ \bibnamefont
			{Long}},\ }\href {\doibase 10.1007/JHEP03(2021)283} {\bibfield  {journal}
		{\bibinfo  {journal} {JHEP}\ }\textbf {\bibinfo {volume} {03}},\ \bibinfo
		{pages} {283} (\bibinfo {year} {2021})},\ \Eprint
	{http://arxiv.org/abs/2009.03828} {arXiv:2009.03828 [astro-ph.CO]}
	\BibitemShut {NoStop}%
	\bibitem [{\citenamefont {Ema}\ \emph {et~al.}(2017)\citenamefont {Ema},
		\citenamefont {Jinno}, \citenamefont {Mukaida},\ and\ \citenamefont
		{Nakayama}}]{Ema:2016dny}%
	\BibitemOpen
	\bibfield  {author} {\bibinfo {author} {\bibfnamefont {Y.}~\bibnamefont
			{Ema}}, \bibinfo {author} {\bibfnamefont {R.}~\bibnamefont {Jinno}}, \bibinfo
		{author} {\bibfnamefont {K.}~\bibnamefont {Mukaida}}, \ and\ \bibinfo
		{author} {\bibfnamefont {K.}~\bibnamefont {Nakayama}},\ }\href {\doibase
		10.1088/1475-7516/2017/02/045} {\bibfield  {journal} {\bibinfo  {journal}
			{JCAP}\ }\textbf {\bibinfo {volume} {02}},\ \bibinfo {pages} {045} (\bibinfo
		{year} {2017})},\ \Eprint {http://arxiv.org/abs/1609.05209} {arXiv:1609.05209
		[hep-ph]} \BibitemShut {NoStop}%
	\bibitem [{\citenamefont {Ema}\ \emph {et~al.}(2016)\citenamefont {Ema},
		\citenamefont {Jinno}, \citenamefont {Mukaida},\ and\ \citenamefont
		{Nakayama}}]{Ema:2016hlw}%
	\BibitemOpen
	\bibfield  {author} {\bibinfo {author} {\bibfnamefont {Y.}~\bibnamefont
			{Ema}}, \bibinfo {author} {\bibfnamefont {R.}~\bibnamefont {Jinno}}, \bibinfo
		{author} {\bibfnamefont {K.}~\bibnamefont {Mukaida}}, \ and\ \bibinfo
		{author} {\bibfnamefont {K.}~\bibnamefont {Nakayama}},\ }\href {\doibase
		10.1103/PhysRevD.94.063517} {\bibfield  {journal} {\bibinfo  {journal} {Phys.
				Rev. D}\ }\textbf {\bibinfo {volume} {94}},\ \bibinfo {pages} {063517}
		(\bibinfo {year} {2016})},\ \Eprint {http://arxiv.org/abs/1604.08898}
	{arXiv:1604.08898 [hep-ph]} \BibitemShut {NoStop}%
	\bibitem [{\citenamefont {Li}\ \emph {et~al.}(2020)\citenamefont {Li},
		\citenamefont {Lu}, \citenamefont {Wang},\ and\ \citenamefont
		{Zhou}}]{Li:2020xwr}%
	\BibitemOpen
	\bibfield  {author} {\bibinfo {author} {\bibfnamefont {L.}~\bibnamefont
			{Li}}, \bibinfo {author} {\bibfnamefont {S.}~\bibnamefont {Lu}}, \bibinfo
		{author} {\bibfnamefont {Y.}~\bibnamefont {Wang}}, \ and\ \bibinfo {author}
		{\bibfnamefont {S.}~\bibnamefont {Zhou}},\ }\href {\doibase
		10.1007/JHEP07(2020)231} {\bibfield  {journal} {\bibinfo  {journal} {JHEP}\
		}\textbf {\bibinfo {volume} {07}},\ \bibinfo {pages} {231} (\bibinfo {year}
		{2020})},\ \Eprint {http://arxiv.org/abs/2002.01131} {arXiv:2002.01131
		[hep-ph]} \BibitemShut {NoStop}%
	\bibitem [{\citenamefont {Li}\ \emph {et~al.}(2019)\citenamefont {Li},
		\citenamefont {Nakama}, \citenamefont {Sou}, \citenamefont {Wang},\ and\
		\citenamefont {Zhou}}]{Li:2019ves}%
	\BibitemOpen
	\bibfield  {author} {\bibinfo {author} {\bibfnamefont {L.}~\bibnamefont
			{Li}}, \bibinfo {author} {\bibfnamefont {T.}~\bibnamefont {Nakama}}, \bibinfo
		{author} {\bibfnamefont {C.~M.}\ \bibnamefont {Sou}}, \bibinfo {author}
		{\bibfnamefont {Y.}~\bibnamefont {Wang}}, \ and\ \bibinfo {author}
		{\bibfnamefont {S.}~\bibnamefont {Zhou}},\ }\href {\doibase
		10.1007/JHEP07(2019)067} {\bibfield  {journal} {\bibinfo  {journal} {JHEP}\
		}\textbf {\bibinfo {volume} {07}},\ \bibinfo {pages} {067} (\bibinfo {year}
		{2019})},\ \Eprint {http://arxiv.org/abs/1903.08842} {arXiv:1903.08842
		[astro-ph.CO]} \BibitemShut {NoStop}%
	\bibitem [{\citenamefont {Herring}\ and\ \citenamefont
		{Boyanovsky}(2020)}]{Herring:2020cah}%
	\BibitemOpen
	\bibfield  {author} {\bibinfo {author} {\bibfnamefont {N.}~\bibnamefont
			{Herring}}\ and\ \bibinfo {author} {\bibfnamefont {D.}~\bibnamefont
			{Boyanovsky}},\ }\href {\doibase 10.1103/PhysRevD.101.123522} {\bibfield
		{journal} {\bibinfo  {journal} {Phys. Rev. D}\ }\textbf {\bibinfo {volume}
			{101}},\ \bibinfo {pages} {123522} (\bibinfo {year} {2020})},\ \Eprint
	{http://arxiv.org/abs/2005.00391} {arXiv:2005.00391 [astro-ph.CO]}
	\BibitemShut {NoStop}%
	\bibitem [{\citenamefont {Babichev}\ \emph {et~al.}(2019)\citenamefont
		{Babichev}, \citenamefont {Gorbunov},\ and\ \citenamefont
		{Ramazanov}}]{Babichev:2018mtd}%
	\BibitemOpen
	\bibfield  {author} {\bibinfo {author} {\bibfnamefont {E.}~\bibnamefont
			{Babichev}}, \bibinfo {author} {\bibfnamefont {D.}~\bibnamefont {Gorbunov}},
		\ and\ \bibinfo {author} {\bibfnamefont {S.}~\bibnamefont {Ramazanov}},\
	}\href {\doibase 10.1016/j.physletb.2019.05.030} {\bibfield  {journal}
		{\bibinfo  {journal} {Phys. Lett. B}\ }\textbf {\bibinfo {volume} {794}},\
		\bibinfo {pages} {69} (\bibinfo {year} {2019})},\ \Eprint
	{http://arxiv.org/abs/1812.03516} {arXiv:1812.03516 [hep-ph]} \BibitemShut
	{NoStop}%
	\bibitem [{\citenamefont {Hashiba}\ and\ \citenamefont
		{Yokoyama}(2019)}]{Hashiba:2018tbu}%
	\BibitemOpen
	\bibfield  {author} {\bibinfo {author} {\bibfnamefont {S.}~\bibnamefont
			{Hashiba}}\ and\ \bibinfo {author} {\bibfnamefont {J.}~\bibnamefont
			{Yokoyama}},\ }\href {\doibase 10.1103/PhysRevD.99.043008} {\bibfield
		{journal} {\bibinfo  {journal} {Phys. Rev. D}\ }\textbf {\bibinfo {volume}
			{99}},\ \bibinfo {pages} {043008} (\bibinfo {year} {2019})},\ \Eprint
	{http://arxiv.org/abs/1812.10032} {arXiv:1812.10032 [hep-ph]} \BibitemShut
	{NoStop}%
	\bibitem [{\citenamefont {Ling}\ and\ \citenamefont
		{Long}(2021)}]{Ling:2021zlj}%
	\BibitemOpen
	\bibfield  {author} {\bibinfo {author} {\bibfnamefont {S.}~\bibnamefont
			{Ling}}\ and\ \bibinfo {author} {\bibfnamefont {A.~J.}\ \bibnamefont
			{Long}},\ }\href {\doibase 10.1103/PhysRevD.103.103532} {\bibfield  {journal}
		{\bibinfo  {journal} {Phys. Rev. D}\ }\textbf {\bibinfo {volume} {103}},\
		\bibinfo {pages} {103532} (\bibinfo {year} {2021})},\ \Eprint
	{http://arxiv.org/abs/2101.11621} {arXiv:2101.11621 [astro-ph.CO]}
	\BibitemShut {NoStop}%
	\bibitem [{\citenamefont {Garny}\ \emph {et~al.}(2016)\citenamefont {Garny},
		\citenamefont {Sandora},\ and\ \citenamefont {Sloth}}]{Garny:2015sjg}%
	\BibitemOpen
	\bibfield  {author} {\bibinfo {author} {\bibfnamefont {M.}~\bibnamefont
			{Garny}}, \bibinfo {author} {\bibfnamefont {M.}~\bibnamefont {Sandora}}, \
		and\ \bibinfo {author} {\bibfnamefont {M.~S.}\ \bibnamefont {Sloth}},\ }\href
	{\doibase 10.1103/PhysRevLett.116.101302} {\bibfield  {journal} {\bibinfo
			{journal} {Phys. Rev. Lett.}\ }\textbf {\bibinfo {volume} {116}},\ \bibinfo
		{pages} {101302} (\bibinfo {year} {2016})},\ \Eprint
	{http://arxiv.org/abs/1511.03278} {arXiv:1511.03278 [hep-ph]} \BibitemShut
	{NoStop}%
	\bibitem [{\citenamefont {Tang}\ and\ \citenamefont {Wu}(2016)}]{Tang:2016vch}%
	\BibitemOpen
	\bibfield  {author} {\bibinfo {author} {\bibfnamefont {Y.}~\bibnamefont
			{Tang}}\ and\ \bibinfo {author} {\bibfnamefont {Y.-L.}\ \bibnamefont {Wu}},\
	}\href {\doibase 10.1016/j.physletb.2016.05.045} {\bibfield  {journal}
		{\bibinfo  {journal} {Phys. Lett. B}\ }\textbf {\bibinfo {volume} {758}},\
		\bibinfo {pages} {402} (\bibinfo {year} {2016})},\ \Eprint
	{http://arxiv.org/abs/1604.04701} {arXiv:1604.04701 [hep-ph]} \BibitemShut
	{NoStop}%
	\bibitem [{\citenamefont {Tang}\ and\ \citenamefont {Wu}(2017)}]{pgdm3}%
	\BibitemOpen
	\bibfield  {author} {\bibinfo {author} {\bibfnamefont {Y.}~\bibnamefont
			{Tang}}\ and\ \bibinfo {author} {\bibfnamefont {Y.-L.}\ \bibnamefont {Wu}},\
	}\href {\doibase 10.1016/j.physletb.2017.10.034} {\bibfield  {journal}
		{\bibinfo  {journal} {Phys. Lett. B}\ }\textbf {\bibinfo {volume} {774}},\
		\bibinfo {pages} {676} (\bibinfo {year} {2017})},\ \Eprint
	{http://arxiv.org/abs/1708.05138} {arXiv:1708.05138 [hep-ph]} \BibitemShut
	{NoStop}%
	\bibitem [{\citenamefont {Garny}\ \emph {et~al.}(2018)\citenamefont {Garny},
		\citenamefont {Palessandro}, \citenamefont {Sandora},\ and\ \citenamefont
		{Sloth}}]{Garny:2017kha}%
	\BibitemOpen
	\bibfield  {author} {\bibinfo {author} {\bibfnamefont {M.}~\bibnamefont
			{Garny}}, \bibinfo {author} {\bibfnamefont {A.}~\bibnamefont {Palessandro}},
		\bibinfo {author} {\bibfnamefont {M.}~\bibnamefont {Sandora}}, \ and\
		\bibinfo {author} {\bibfnamefont {M.~S.}\ \bibnamefont {Sloth}},\ }\href
	{\doibase 10.1088/1475-7516/2018/02/027} {\bibfield  {journal} {\bibinfo
			{journal} {JCAP}\ }\textbf {\bibinfo {volume} {02}},\ \bibinfo {pages} {027}
		(\bibinfo {year} {2018})},\ \Eprint {http://arxiv.org/abs/1709.09688}
	{arXiv:1709.09688 [hep-ph]} \BibitemShut {NoStop}%
	\bibitem [{\citenamefont {Chen}\ and\ \citenamefont
		{Kang}(2018)}]{Chen:2017kvz}%
	\BibitemOpen
	\bibfield  {author} {\bibinfo {author} {\bibfnamefont {S.-L.}\ \bibnamefont
			{Chen}}\ and\ \bibinfo {author} {\bibfnamefont {Z.}~\bibnamefont {Kang}},\
	}\href {\doibase 10.1088/1475-7516/2018/05/036} {\bibfield  {journal}
		{\bibinfo  {journal} {JCAP}\ }\textbf {\bibinfo {volume} {05}},\ \bibinfo
		{pages} {036} (\bibinfo {year} {2018})},\ \Eprint
	{http://arxiv.org/abs/1711.02556} {arXiv:1711.02556 [hep-ph]} \BibitemShut
	{NoStop}%
	\bibitem [{\citenamefont {Bernal}\ \emph {et~al.}(2018)\citenamefont {Bernal},
		\citenamefont {Dutra}, \citenamefont {Mambrini}, \citenamefont {Olive},
		\citenamefont {Peloso},\ and\ \citenamefont {Pierre}}]{Bernal:2018qlk}%
	\BibitemOpen
	\bibfield  {author} {\bibinfo {author} {\bibfnamefont {N.}~\bibnamefont
			{Bernal}}, \bibinfo {author} {\bibfnamefont {M.}~\bibnamefont {Dutra}},
		\bibinfo {author} {\bibfnamefont {Y.}~\bibnamefont {Mambrini}}, \bibinfo
		{author} {\bibfnamefont {K.}~\bibnamefont {Olive}}, \bibinfo {author}
		{\bibfnamefont {M.}~\bibnamefont {Peloso}}, \ and\ \bibinfo {author}
		{\bibfnamefont {M.}~\bibnamefont {Pierre}},\ }\href {\doibase
		10.1103/PhysRevD.97.115020} {\bibfield  {journal} {\bibinfo  {journal} {Phys.
				Rev. D}\ }\textbf {\bibinfo {volume} {97}},\ \bibinfo {pages} {115020}
		(\bibinfo {year} {2018})},\ \Eprint {http://arxiv.org/abs/1803.01866}
	{arXiv:1803.01866 [hep-ph]} \BibitemShut {NoStop}%
	\bibitem [{\citenamefont {Aoki}\ \emph {et~al.}(2022)\citenamefont {Aoki},
		\citenamefont {Lee}, \citenamefont {Menkara},\ and\ \citenamefont
		{Yamashita}}]{Aoki:2022dzd}%
	\BibitemOpen
	\bibfield  {author} {\bibinfo {author} {\bibfnamefont {S.}~\bibnamefont
			{Aoki}}, \bibinfo {author} {\bibfnamefont {H.~M.}\ \bibnamefont {Lee}},
		\bibinfo {author} {\bibfnamefont {A.~G.}\ \bibnamefont {Menkara}}, \ and\
		\bibinfo {author} {\bibfnamefont {K.}~\bibnamefont {Yamashita}},\ }\href@noop
	{} {\  (\bibinfo {year} {2022})},\ \Eprint {http://arxiv.org/abs/2202.13063}
	{arXiv:2202.13063 [hep-ph]} \BibitemShut {NoStop}%
	\bibitem [{\citenamefont {Clery}\ \emph {et~al.}(2021)\citenamefont {Clery},
		\citenamefont {Mambrini}, \citenamefont {Olive},\ and\ \citenamefont
		{Verner}}]{Clery:2021bwz}%
	\BibitemOpen
	\bibfield  {author} {\bibinfo {author} {\bibfnamefont {S.}~\bibnamefont
			{Clery}}, \bibinfo {author} {\bibfnamefont {Y.}~\bibnamefont {Mambrini}},
		\bibinfo {author} {\bibfnamefont {K.~A.}\ \bibnamefont {Olive}}, \ and\
		\bibinfo {author} {\bibfnamefont {S.}~\bibnamefont {Verner}},\ }\href@noop {}
	{\  (\bibinfo {year} {2021})},\ \Eprint {http://arxiv.org/abs/2112.15214}
	{arXiv:2112.15214 [hep-ph]} \BibitemShut {NoStop}%
	\bibitem [{\citenamefont {Frangipane}\ \emph {et~al.}(2021)\citenamefont
		{Frangipane}, \citenamefont {Gori},\ and\ \citenamefont
		{Shakya}}]{Frangipane:2021rtf}%
	\BibitemOpen
	\bibfield  {author} {\bibinfo {author} {\bibfnamefont {E.}~\bibnamefont
			{Frangipane}}, \bibinfo {author} {\bibfnamefont {S.}~\bibnamefont {Gori}}, \
		and\ \bibinfo {author} {\bibfnamefont {B.}~\bibnamefont {Shakya}},\
	}\href@noop {} {\  (\bibinfo {year} {2021})},\ \Eprint
	{http://arxiv.org/abs/2110.10711} {arXiv:2110.10711 [hep-ph]} \BibitemShut
	{NoStop}%
	\bibitem [{\citenamefont {Redi}\ and\ \citenamefont
		{Tesi}(2021)}]{Redi:2021ipn}%
	\BibitemOpen
	\bibfield  {author} {\bibinfo {author} {\bibfnamefont {M.}~\bibnamefont
			{Redi}}\ and\ \bibinfo {author} {\bibfnamefont {A.}~\bibnamefont {Tesi}},\
	}\href {\doibase 10.1007/JHEP12(2021)060} {\bibfield  {journal} {\bibinfo
			{journal} {JHEP}\ }\textbf {\bibinfo {volume} {12}},\ \bibinfo {pages} {060}
		(\bibinfo {year} {2021})},\ \Eprint {http://arxiv.org/abs/2107.14801}
	{arXiv:2107.14801 [hep-ph]} \BibitemShut {NoStop}%
	\bibitem [{\citenamefont {Barman}\ \emph {et~al.}(2022)\citenamefont {Barman},
		\citenamefont {Bernal}, \citenamefont {Das},\ and\ \citenamefont
		{Roshan}}]{Barman:2021qds}%
	\BibitemOpen
	\bibfield  {author} {\bibinfo {author} {\bibfnamefont {B.}~\bibnamefont
			{Barman}}, \bibinfo {author} {\bibfnamefont {N.}~\bibnamefont {Bernal}},
		\bibinfo {author} {\bibfnamefont {A.}~\bibnamefont {Das}}, \ and\ \bibinfo
		{author} {\bibfnamefont {R.}~\bibnamefont {Roshan}},\ }\href {\doibase
		10.1088/1475-7516/2022/01/047} {\bibfield  {journal} {\bibinfo  {journal}
			{JCAP}\ }\textbf {\bibinfo {volume} {01}},\ \bibinfo {pages} {047} (\bibinfo
		{year} {2022})},\ \Eprint {http://arxiv.org/abs/2108.13447} {arXiv:2108.13447
		[hep-ph]} \BibitemShut {NoStop}%
	\bibitem [{\citenamefont {Mambrini}\ and\ \citenamefont
		{Olive}(2021)}]{Mambrini:2021zpp}%
	\BibitemOpen
	\bibfield  {author} {\bibinfo {author} {\bibfnamefont {Y.}~\bibnamefont
			{Mambrini}}\ and\ \bibinfo {author} {\bibfnamefont {K.~A.}\ \bibnamefont
			{Olive}},\ }\href {\doibase 10.1103/PhysRevD.103.115009} {\bibfield
		{journal} {\bibinfo  {journal} {Phys. Rev. D}\ }\textbf {\bibinfo {volume}
			{103}},\ \bibinfo {pages} {115009} (\bibinfo {year} {2021})},\ \Eprint
	{http://arxiv.org/abs/2102.06214} {arXiv:2102.06214 [hep-ph]} \BibitemShut
	{NoStop}%
	\bibitem [{\citenamefont {Haque}\ and\ \citenamefont
		{Maity}(2021)}]{Haque:2021mab}%
	\BibitemOpen
	\bibfield  {author} {\bibinfo {author} {\bibfnamefont {M.~R.}\ \bibnamefont
			{Haque}}\ and\ \bibinfo {author} {\bibfnamefont {D.}~\bibnamefont {Maity}},\
	}\href@noop {} {\  (\bibinfo {year} {2021})},\ \Eprint
	{http://arxiv.org/abs/2112.14668} {arXiv:2112.14668 [hep-ph]} \BibitemShut
	{NoStop}%
	\bibitem [{\citenamefont {Clery}\ \emph {et~al.}(2022)\citenamefont {Clery},
		\citenamefont {Mambrini}, \citenamefont {Olive}, \citenamefont {Shkerin},\
		and\ \citenamefont {Verner}}]{Clery:2022wib}%
	\BibitemOpen
	\bibfield  {author} {\bibinfo {author} {\bibfnamefont {S.}~\bibnamefont
			{Clery}}, \bibinfo {author} {\bibfnamefont {Y.}~\bibnamefont {Mambrini}},
		\bibinfo {author} {\bibfnamefont {K.~A.}\ \bibnamefont {Olive}}, \bibinfo
		{author} {\bibfnamefont {A.}~\bibnamefont {Shkerin}}, \ and\ \bibinfo
		{author} {\bibfnamefont {S.}~\bibnamefont {Verner}},\ }\href@noop {} {\
		(\bibinfo {year} {2022})},\ \Eprint {http://arxiv.org/abs/2203.02004}
	{arXiv:2203.02004 [hep-ph]} \BibitemShut {NoStop}%
	\bibitem [{\citenamefont {Wu}(2021{\natexlab{a}})}]{Wu:2021ign}%
	\BibitemOpen
	\bibfield  {author} {\bibinfo {author} {\bibfnamefont {Y.-L.}\ \bibnamefont
			{Wu}},\ }\href {\doibase 10.1142/S0217751X21430016} {\bibfield  {journal}
		{\bibinfo  {journal} {Int. J. Mod. Phys. A}\ }\textbf {\bibinfo {volume}
			{36}},\ \bibinfo {pages} {2143001} (\bibinfo {year} {2021}{\natexlab{a}})},\
	\Eprint {http://arxiv.org/abs/2104.05404} {arXiv:2104.05404 [physics.gen-ph]}
	\BibitemShut {NoStop}%
	\bibitem [{\citenamefont {Wu}(2021{\natexlab{b}})}]{Wu:2021ucc}%
	\BibitemOpen
	\bibfield  {author} {\bibinfo {author} {\bibfnamefont {Y.-L.}\ \bibnamefont
			{Wu}},\ }\href {\doibase 10.1142/S0217751X21430028} {\bibfield  {journal}
		{\bibinfo  {journal} {Int. J. Mod. Phys. A}\ }\textbf {\bibinfo {volume}
			{36}},\ \bibinfo {pages} {2143002} (\bibinfo {year} {2021}{\natexlab{b}})},\
	\Eprint {http://arxiv.org/abs/2104.11078} {arXiv:2104.11078 [physics.gen-ph]}
	\BibitemShut {NoStop}%
	\bibitem [{\citenamefont {Barrow}\ and\ \citenamefont {Cotsakis}(1988)}]{fR1}%
	\BibitemOpen
	\bibfield  {author} {\bibinfo {author} {\bibfnamefont {J.~D.}\ \bibnamefont
			{Barrow}}\ and\ \bibinfo {author} {\bibfnamefont {S.}~\bibnamefont
			{Cotsakis}},\ }\href {\doibase 10.1016/0370-2693(88)90110-4} {\bibfield
		{journal} {\bibinfo  {journal} {Phys. Lett. B}\ }\textbf {\bibinfo {volume}
			{214}},\ \bibinfo {pages} {515} (\bibinfo {year} {1988})}\BibitemShut
	{NoStop}%
	\bibitem [{\citenamefont {De~Felice}\ and\ \citenamefont
		{Tsujikawa}(2010)}]{fR2}%
	\BibitemOpen
	\bibfield  {author} {\bibinfo {author} {\bibfnamefont {A.}~\bibnamefont
			{De~Felice}}\ and\ \bibinfo {author} {\bibfnamefont {S.}~\bibnamefont
			{Tsujikawa}},\ }\href {\doibase 10.12942/lrr-2010-3} {\bibfield  {journal}
		{\bibinfo  {journal} {Living Rev. Rel.}\ }\textbf {\bibinfo {volume} {13}},\
		\bibinfo {pages} {3} (\bibinfo {year} {2010})},\ \Eprint
	{http://arxiv.org/abs/1002.4928} {arXiv:1002.4928 [gr-qc]} \BibitemShut
	{NoStop}%
	\bibitem [{\citenamefont {Starobinsky}(1980)}]{Staro1}%
	\BibitemOpen
	\bibfield  {author} {\bibinfo {author} {\bibfnamefont {A.~A.}\ \bibnamefont
			{Starobinsky}},\ }\href {\doibase 10.1016/0370-2693(80)90670-X} {\bibfield
		{journal} {\bibinfo  {journal} {Phys. Lett. B}\ }\textbf {\bibinfo {volume}
			{91}},\ \bibinfo {pages} {99} (\bibinfo {year} {1980})}\BibitemShut {NoStop}%
	\bibitem [{\citenamefont {Maeda}(1988)}]{Staro2}%
	\BibitemOpen
	\bibfield  {author} {\bibinfo {author} {\bibfnamefont {K.-i.}\ \bibnamefont
			{Maeda}},\ }\href {\doibase 10.1103/PhysRevD.37.858} {\bibfield  {journal}
		{\bibinfo  {journal} {Phys. Rev. D}\ }\textbf {\bibinfo {volume} {37}},\
		\bibinfo {pages} {858} (\bibinfo {year} {1988})}\BibitemShut {NoStop}%
	\bibitem [{\citenamefont {Cembranos}(2009)}]{Staro3}%
	\BibitemOpen
	\bibfield  {author} {\bibinfo {author} {\bibfnamefont {J.~A.~R.}\
			\bibnamefont {Cembranos}},\ }\href {\doibase 10.1103/PhysRevLett.102.141301}
	{\bibfield  {journal} {\bibinfo  {journal} {Phys. Rev. Lett.}\ }\textbf
		{\bibinfo {volume} {102}},\ \bibinfo {pages} {141301} (\bibinfo {year}
		{2009})},\ \Eprint {http://arxiv.org/abs/0809.1653} {arXiv:0809.1653
		[hep-ph]} \BibitemShut {NoStop}%
	\bibitem [{\citenamefont {Bezrukov}\ and\ \citenamefont
		{Gorbunov}(2012)}]{Staro4}%
	\BibitemOpen
	\bibfield  {author} {\bibinfo {author} {\bibfnamefont {F.~L.}\ \bibnamefont
			{Bezrukov}}\ and\ \bibinfo {author} {\bibfnamefont {D.~S.}\ \bibnamefont
			{Gorbunov}},\ }\href {\doibase 10.1016/j.physletb.2012.06.040} {\bibfield
		{journal} {\bibinfo  {journal} {Phys. Lett. B}\ }\textbf {\bibinfo {volume}
			{713}},\ \bibinfo {pages} {365} (\bibinfo {year} {2012})},\ \Eprint
	{http://arxiv.org/abs/1111.4397} {arXiv:1111.4397 [hep-ph]} \BibitemShut
	{NoStop}%
	\bibitem [{\citenamefont {He}\ \emph {et~al.}(2018)\citenamefont {He},
		\citenamefont {Starobinsky},\ and\ \citenamefont {Yokoyama}}]{Staro5}%
	\BibitemOpen
	\bibfield  {author} {\bibinfo {author} {\bibfnamefont {M.}~\bibnamefont
			{He}}, \bibinfo {author} {\bibfnamefont {A.~A.}\ \bibnamefont {Starobinsky}},
		\ and\ \bibinfo {author} {\bibfnamefont {J.}~\bibnamefont {Yokoyama}},\
	}\href {\doibase 10.1088/1475-7516/2018/05/064} {\bibfield  {journal}
		{\bibinfo  {journal} {JCAP}\ }\textbf {\bibinfo {volume} {05}},\ \bibinfo
		{pages} {064} (\bibinfo {year} {2018})},\ \Eprint
	{http://arxiv.org/abs/1804.00409} {arXiv:1804.00409 [astro-ph.CO]}
	\BibitemShut {NoStop}%
	\bibitem [{\citenamefont {Enckell}\ \emph {et~al.}(2019)\citenamefont
		{Enckell}, \citenamefont {Enqvist}, \citenamefont {Rasanen},\ and\
		\citenamefont {Wahlman}}]{Staro6}%
	\BibitemOpen
	\bibfield  {author} {\bibinfo {author} {\bibfnamefont {V.-M.}\ \bibnamefont
			{Enckell}}, \bibinfo {author} {\bibfnamefont {K.}~\bibnamefont {Enqvist}},
		\bibinfo {author} {\bibfnamefont {S.}~\bibnamefont {Rasanen}}, \ and\
		\bibinfo {author} {\bibfnamefont {L.-P.}\ \bibnamefont {Wahlman}},\ }\href
	{\doibase 10.1088/1475-7516/2019/02/022} {\bibfield  {journal} {\bibinfo
			{journal} {JCAP}\ }\textbf {\bibinfo {volume} {02}},\ \bibinfo {pages} {022}
		(\bibinfo {year} {2019})},\ \Eprint {http://arxiv.org/abs/1810.05536}
	{arXiv:1810.05536 [gr-qc]} \BibitemShut {NoStop}%
	\bibitem [{\citenamefont {Pi}\ \emph {et~al.}(2018)\citenamefont {Pi},
		\citenamefont {Zhang}, \citenamefont {Huang},\ and\ \citenamefont
		{Sasaki}}]{Staro7}%
	\BibitemOpen
	\bibfield  {author} {\bibinfo {author} {\bibfnamefont {S.}~\bibnamefont
			{Pi}}, \bibinfo {author} {\bibfnamefont {Y.-l.}\ \bibnamefont {Zhang}},
		\bibinfo {author} {\bibfnamefont {Q.-G.}\ \bibnamefont {Huang}}, \ and\
		\bibinfo {author} {\bibfnamefont {M.}~\bibnamefont {Sasaki}},\ }\href
	{\doibase 10.1088/1475-7516/2018/05/042} {\bibfield  {journal} {\bibinfo
			{journal} {JCAP}\ }\textbf {\bibinfo {volume} {05}},\ \bibinfo {pages} {042}
		(\bibinfo {year} {2018})},\ \Eprint {http://arxiv.org/abs/1712.09896}
	{arXiv:1712.09896 [astro-ph.CO]} \BibitemShut {NoStop}%
	\bibitem [{\citenamefont {Antoniadis}\ \emph {et~al.}(2018)\citenamefont
		{Antoniadis}, \citenamefont {Karam}, \citenamefont {Lykkas},\ and\
		\citenamefont {Tamvakis}}]{Staro8}%
	\BibitemOpen
	\bibfield  {author} {\bibinfo {author} {\bibfnamefont {I.}~\bibnamefont
			{Antoniadis}}, \bibinfo {author} {\bibfnamefont {A.}~\bibnamefont {Karam}},
		\bibinfo {author} {\bibfnamefont {A.}~\bibnamefont {Lykkas}}, \ and\ \bibinfo
		{author} {\bibfnamefont {K.}~\bibnamefont {Tamvakis}},\ }\href {\doibase
		10.1088/1475-7516/2018/11/028} {\bibfield  {journal} {\bibinfo  {journal}
			{JCAP}\ }\textbf {\bibinfo {volume} {11}},\ \bibinfo {pages} {028} (\bibinfo
		{year} {2018})},\ \Eprint {http://arxiv.org/abs/1810.10418} {arXiv:1810.10418
		[gr-qc]} \BibitemShut {NoStop}%
	\bibitem [{\citenamefont {Ade}\ \emph {et~al.}(2021)\citenamefont {Ade} \emph
		{et~al.}}]{BICEP}%
	\BibitemOpen
	\bibfield  {author} {\bibinfo {author} {\bibfnamefont {P.~A.~R.}\
			\bibnamefont {Ade}} \emph {et~al.} (\bibinfo {collaboration} {BICEP, Keck}),\
	}\href {\doibase 10.1103/PhysRevLett.127.151301} {\bibfield  {journal}
		{\bibinfo  {journal} {Phys. Rev. Lett.}\ }\textbf {\bibinfo {volume} {127}},\
		\bibinfo {pages} {151301} (\bibinfo {year} {2021})},\ \Eprint
	{http://arxiv.org/abs/2110.00483} {arXiv:2110.00483 [astro-ph.CO]}
	\BibitemShut {NoStop}%
	\bibitem [{\citenamefont {Liddle}\ and\ \citenamefont {Leach}(2003)}]{efolds}%
	\BibitemOpen
	\bibfield  {author} {\bibinfo {author} {\bibfnamefont {A.~R.}\ \bibnamefont
			{Liddle}}\ and\ \bibinfo {author} {\bibfnamefont {S.~M.}\ \bibnamefont
			{Leach}},\ }\href {\doibase 10.1103/PhysRevD.68.103503} {\bibfield  {journal}
		{\bibinfo  {journal} {Phys. Rev. D}\ }\textbf {\bibinfo {volume} {68}},\
		\bibinfo {pages} {103503} (\bibinfo {year} {2003})},\ \Eprint
	{http://arxiv.org/abs/astro-ph/0305263} {arXiv:astro-ph/0305263} \BibitemShut
	{NoStop}%
	\bibitem [{\citenamefont {Abazajian}\ \emph {et~al.}(2022)\citenamefont
		{Abazajian} \emph {et~al.}}]{CMB-S4}%
	\BibitemOpen
	\bibfield  {author} {\bibinfo {author} {\bibfnamefont {K.}~\bibnamefont
			{Abazajian}} \emph {et~al.} (\bibinfo {collaboration} {CMB-S4}),\ }\href
	{\doibase 10.3847/1538-4357/ac1596} {\bibfield  {journal} {\bibinfo
			{journal} {Astrophys. J.}\ }\textbf {\bibinfo {volume} {926}},\ \bibinfo
		{pages} {54} (\bibinfo {year} {2022})},\ \Eprint
	{http://arxiv.org/abs/2008.12619} {arXiv:2008.12619 [astro-ph.CO]}
	\BibitemShut {NoStop}%
	\bibitem [{\citenamefont {Kofman}\ \emph {et~al.}(1997)\citenamefont {Kofman},
		\citenamefont {Linde},\ and\ \citenamefont {Starobinsky}}]{reh1}%
	\BibitemOpen
	\bibfield  {author} {\bibinfo {author} {\bibfnamefont {L.}~\bibnamefont
			{Kofman}}, \bibinfo {author} {\bibfnamefont {A.~D.}\ \bibnamefont {Linde}}, \
		and\ \bibinfo {author} {\bibfnamefont {A.~A.}\ \bibnamefont {Starobinsky}},\
	}\href {\doibase 10.1103/PhysRevD.56.3258} {\bibfield  {journal} {\bibinfo
			{journal} {Phys. Rev. D}\ }\textbf {\bibinfo {volume} {56}},\ \bibinfo
		{pages} {3258} (\bibinfo {year} {1997})},\ \Eprint
	{http://arxiv.org/abs/hep-ph/9704452} {arXiv:hep-ph/9704452} \BibitemShut
	{NoStop}%
	\bibitem [{\citenamefont {Albrecht}\ \emph {et~al.}(1982)\citenamefont
		{Albrecht}, \citenamefont {Steinhardt}, \citenamefont {Turner},\ and\
		\citenamefont {Wilczek}}]{reh2}%
	\BibitemOpen
	\bibfield  {author} {\bibinfo {author} {\bibfnamefont {A.}~\bibnamefont
			{Albrecht}}, \bibinfo {author} {\bibfnamefont {P.~J.}\ \bibnamefont
			{Steinhardt}}, \bibinfo {author} {\bibfnamefont {M.~S.}\ \bibnamefont
			{Turner}}, \ and\ \bibinfo {author} {\bibfnamefont {F.}~\bibnamefont
			{Wilczek}},\ }\href {\doibase 10.1103/PhysRevLett.48.1437} {\bibfield
		{journal} {\bibinfo  {journal} {Phys. Rev. Lett.}\ }\textbf {\bibinfo
			{volume} {48}},\ \bibinfo {pages} {1437} (\bibinfo {year}
		{1982})}\BibitemShut {NoStop}%
	\bibitem [{\citenamefont {Kofman}\ \emph {et~al.}(1994)\citenamefont {Kofman},
		\citenamefont {Linde},\ and\ \citenamefont {Starobinsky}}]{reh3}%
	\BibitemOpen
	\bibfield  {author} {\bibinfo {author} {\bibfnamefont {L.}~\bibnamefont
			{Kofman}}, \bibinfo {author} {\bibfnamefont {A.~D.}\ \bibnamefont {Linde}}, \
		and\ \bibinfo {author} {\bibfnamefont {A.~A.}\ \bibnamefont {Starobinsky}},\
	}\href {\doibase 10.1103/PhysRevLett.73.3195} {\bibfield  {journal} {\bibinfo
			{journal} {Phys. Rev. Lett.}\ }\textbf {\bibinfo {volume} {73}},\ \bibinfo
		{pages} {3195} (\bibinfo {year} {1994})},\ \Eprint
	{http://arxiv.org/abs/hep-th/9405187} {arXiv:hep-th/9405187} \BibitemShut
	{NoStop}%
	\bibitem [{\citenamefont {Bassett}\ \emph {et~al.}(2006)\citenamefont
		{Bassett}, \citenamefont {Tsujikawa},\ and\ \citenamefont {Wands}}]{reh4}%
	\BibitemOpen
	\bibfield  {author} {\bibinfo {author} {\bibfnamefont {B.~A.}\ \bibnamefont
			{Bassett}}, \bibinfo {author} {\bibfnamefont {S.}~\bibnamefont {Tsujikawa}},
		\ and\ \bibinfo {author} {\bibfnamefont {D.}~\bibnamefont {Wands}},\ }\href
	{\doibase 10.1103/RevModPhys.78.537} {\bibfield  {journal} {\bibinfo
			{journal} {Rev. Mod. Phys.}\ }\textbf {\bibinfo {volume} {78}},\ \bibinfo
		{pages} {537} (\bibinfo {year} {2006})},\ \Eprint
	{http://arxiv.org/abs/astro-ph/0507632} {arXiv:astro-ph/0507632} \BibitemShut
	{NoStop}%
	\bibitem [{\citenamefont {Lozanov}(2019)}]{reh5}%
	\BibitemOpen
	\bibfield  {author} {\bibinfo {author} {\bibfnamefont {K.~D.}\ \bibnamefont
			{Lozanov}},\ }\href@noop {} {\  (\bibinfo {year} {2019})},\ \Eprint
	{http://arxiv.org/abs/1907.04402} {arXiv:1907.04402 [astro-ph.CO]}
	\BibitemShut {NoStop}%
	\bibitem [{\citenamefont {Bunch}\ and\ \citenamefont
		{Davies}(1978)}]{BDvacuum}%
	\BibitemOpen
	\bibfield  {author} {\bibinfo {author} {\bibfnamefont {T.~S.}\ \bibnamefont
			{Bunch}}\ and\ \bibinfo {author} {\bibfnamefont {P.~C.~W.}\ \bibnamefont
			{Davies}},\ }\href {\doibase 10.1098/rspa.1978.0060} {\bibfield  {journal}
		{\bibinfo  {journal} {Proc. Roy. Soc. Lond. A}\ }\textbf {\bibinfo {volume}
			{360}},\ \bibinfo {pages} {117} (\bibinfo {year} {1978})}\BibitemShut
	{NoStop}%
	\bibitem [{\citenamefont {Hall}\ \emph {et~al.}(2010)\citenamefont {Hall},
		\citenamefont {Jedamzik}, \citenamefont {March-Russell},\ and\ \citenamefont
		{West}}]{freezein1}%
	\BibitemOpen
	\bibfield  {author} {\bibinfo {author} {\bibfnamefont {L.~J.}\ \bibnamefont
			{Hall}}, \bibinfo {author} {\bibfnamefont {K.}~\bibnamefont {Jedamzik}},
		\bibinfo {author} {\bibfnamefont {J.}~\bibnamefont {March-Russell}}, \ and\
		\bibinfo {author} {\bibfnamefont {S.~M.}\ \bibnamefont {West}},\ }\href
	{\doibase 10.1007/JHEP03(2010)080} {\bibfield  {journal} {\bibinfo  {journal}
			{JHEP}\ }\textbf {\bibinfo {volume} {03}},\ \bibinfo {pages} {080} (\bibinfo
		{year} {2010})},\ \Eprint {http://arxiv.org/abs/0911.1120} {arXiv:0911.1120
		[hep-ph]} \BibitemShut {NoStop}%
	\bibitem [{\citenamefont {Blennow}\ \emph {et~al.}(2014)\citenamefont
		{Blennow}, \citenamefont {Fernandez-Martinez},\ and\ \citenamefont
		{Zaldivar}}]{freezein2}%
	\BibitemOpen
	\bibfield  {author} {\bibinfo {author} {\bibfnamefont {M.}~\bibnamefont
			{Blennow}}, \bibinfo {author} {\bibfnamefont {E.}~\bibnamefont
			{Fernandez-Martinez}}, \ and\ \bibinfo {author} {\bibfnamefont
			{B.}~\bibnamefont {Zaldivar}},\ }\href {\doibase
		10.1088/1475-7516/2014/01/003} {\bibfield  {journal} {\bibinfo  {journal}
			{JCAP}\ }\textbf {\bibinfo {volume} {01}},\ \bibinfo {pages} {003} (\bibinfo
		{year} {2014})},\ \Eprint {http://arxiv.org/abs/1309.7348} {arXiv:1309.7348
		[hep-ph]} \BibitemShut {NoStop}%
	\bibitem [{\citenamefont {Bernal}\ \emph {et~al.}(2017)\citenamefont {Bernal},
		\citenamefont {Heikinheimo}, \citenamefont {Tenkanen}, \citenamefont
		{Tuominen},\ and\ \citenamefont {Vaskonen}}]{FIMP}%
	\BibitemOpen
	\bibfield  {author} {\bibinfo {author} {\bibfnamefont {N.}~\bibnamefont
			{Bernal}}, \bibinfo {author} {\bibfnamefont {M.}~\bibnamefont {Heikinheimo}},
		\bibinfo {author} {\bibfnamefont {T.}~\bibnamefont {Tenkanen}}, \bibinfo
		{author} {\bibfnamefont {K.}~\bibnamefont {Tuominen}}, \ and\ \bibinfo
		{author} {\bibfnamefont {V.}~\bibnamefont {Vaskonen}},\ }\href {\doibase
		10.1142/S0217751X1730023X} {\bibfield  {journal} {\bibinfo  {journal} {Int.
				J. Mod. Phys. A}\ }\textbf {\bibinfo {volume} {32}},\ \bibinfo {pages}
		{1730023} (\bibinfo {year} {2017})},\ \Eprint
	{http://arxiv.org/abs/1706.07442} {arXiv:1706.07442 [hep-ph]} \BibitemShut
	{NoStop}%
\end{thebibliography}
%

\end{document}